\documentclass[aps,prl,preprintnumbers,showpacs,nofootinbib]{revtex4}
\usepackage[utf8]{inputenc}
\usepackage[english]{babel}
\usepackage{amsmath}
\usepackage{amsfonts}
\usepackage{amssymb}
\usepackage{epsfig}
\usepackage{graphics,psfrag,rotating}
\usepackage{graphicx}
\usepackage{dcolumn}
\usepackage{bm}
\usepackage{epstopdf}
\usepackage{color}
\usepackage[usenames,dvipsnames,svgnames]{xcolor}
\usepackage[colorlinks=true,
            linkcolor=red,
            urlcolor=gray,
            citecolor=blue]{hyperref}
  \usepackage{hyperref}

\newcommand{\lsim}   {\mathrel{\mathop{\kern 0pt \rlap
{\raise.2ex\hbox{$<$}}}
 \lower.9ex\hbox{\kern-.190em $\sim$}}}
\newcommand{\gsim}   {\mathrel{\mathop{\kern 0pt \rlap
{\raise.2ex\hbox{$>$}}}
\lower.9ex\hbox{\kern-.190em $\sim$}}}

\def\3nab{\tilde{\nabla}}

\def\hsp5{\hspace{5mm}}

\def\case#1/#2{\textstyle\frac{#1}{#2}}

\def\ber {\begin{eqnarray}}
\def\eer {\end{eqnarray}}
\def\bea {\begin{eqnarray}}
\def\eea {\end{eqnarray}}

\def\bc {\begin{center}}
\def\ec {\end{center}}
\def\case#1/#2{\frac{#1}{#2}}

\newcommand{\bw}{\begin{widetext}}
\newcommand{\ew}{\end{widetext}}

\newcommand{\be}{\begin{equation}}
\newcommand{\bse}{\begin{subequation}}
\newcommand{\ese}{\end{subequation}}
\newcommand{\ee}{\end{equation}}
\newcommand{\eei}{\end{eqnarray}\indent\indent}
\newcommand{\ba}{\begin{array}}
\newcommand{\ea}{\end{array}}
\newcommand{\bal}{\begin{eqnarray}}
\newcommand{\eal}{\end{eqnarray}}

\def\case#1/#2{\textstyle\frac{#1}{#2} }

\newcommand{\bver}{\begin{verbatim}}
\newcommand{\ever}{\end{verbatim}}

\begin{document}


\title{  Barrow Interacting  holographic dark energy cosmology with Hubble horizon as IR cutoff: A model can Alleviating the Hubble and $S_{8}$ Tension}
\author{ Muhammad Yarahmadi$^1$,}
\affiliation{$^{1}$Department of Physics, Lorestan university, Khoramabad, Iran}
\email{yarahmadimohammad10@gmail.com}

\date{\today}

\begin{abstract}
In this study, we perform a comprehensive analysis of the Hubble constant (\(H_0\)) and matter clustering (\(S_8\)) tensions within the framework of non-interacting and interacting Barrow Holographic Dark Energy (BHDE) models. Utilizing a combination of observational datasets, including the Cosmic Microwave Background (CMB), Baryon Acoustic Oscillations (BAO), cosmic chronometers (CC), Pantheon, and lensing data, we assess the degree of tension relative to the Planck 2018 results and recent measurements such as the Riess et al. 2022 (R22) value for \(H_0 = 73.04 \pm 1.04 \ \text{km s}^{-1} \text{Mpc}^{-1}\) in 68\% C.L  and KiDS-1000 and DES-Y3 for \(S_8\). Our findings show that both BHDE models mitigate the \(H_0\) and \(S_8\) tensions compared to the standard \(\Lambda\) Cold Dark Matter (\(\Lambda\)CDM) model. The non-interacting BHDE model achieves a moderate reduction in the \(H_0\) tension, while the interacting BHDE model offers a better fit for both parameters, suggesting it is more effective in addressing the tensions. Additionally, the quantum-gravitational deformation parameter \(\Delta\), constrained using the CMB+All dataset, indicates significant quantum effects in both models. The interacting scenario provides tighter constraints on \(\Delta\) and total neutrino mass \(\sum m_{\nu}\), offering a more precise representation of these effects. This study highlights the potential of BHDE models as viable alternatives to the \(\Lambda\)CDM framework for resolving cosmological tensions.
\end{abstract}

\pacs{}

\keywords{neutrino mass; BHDE; FLRW Universe}

\maketitle

\section{Introduction\label{Int}}
The observational data, derived from various sources including the Cosmic Microwave Background (CMB) \cite{Dunkley}, \cite{Komatsu}, Supernova type Ia (SNIa) \cite{Knop}, \cite{Riess}, Weak Lensing \cite{Leauthand}, Baryon Acoustic Oscillations (BAO) \cite{Parkinson}, the 2dF Galaxy Redshift Survey (2dFGRS) \cite{Cole} at low redshifts, and the DEEP2 redshift survey \cite{Yan} at high redshifts, consistently demonstrate that the universe is predominantly governed by dark energy (DE) with negative pressure. This negative pressure is the driving force behind the universe's accelerated expansion.
In the framework of the Standard Model of particle physics, neutrinos were originally assumed to be massless particles, a property that aligned with the model's initial predictions. In this model, neutrinos are associated with three distinct flavors—electron, muon, and tau—each corresponding to their respective charged leptons. The absence of mass was an essential feature of the Standard Model, as it helped maintain the theory's consistency and symmetry, particularly within the electroweak interaction.

However, more recent experimental evidence has challenged this assumption. It is now known that neutrinos possess three discrete mass eigenstates, each with a tiny but non-zero value. These masses do not align directly with the neutrino flavors. Instead, a neutrino produced with a specific flavor exists as a quantum superposition of all three mass states. This discovery significantly altered our understanding of neutrinos within the Standard Model.

Moreover, experiments on solar and atmospheric neutrino oscillations have provided robust evidence that neutrinos are indeed massive and that there is substantial mixing between different neutrino species. These oscillations, where neutrinos switch between flavors as they travel, can only occur if neutrinos have mass, necessitating a revision of the Standard Model to accommodate this new understanding. This breakthrough has profound implications for particle physics and cosmology, highlighting the need for extensions to the Standard Model to account for the mass and mixing phenomena observed in neutrinos \cite{Lesgourgues}, \cite{Z. z. Xing}.

Recently, several studies have focused on constraining the total neutrino mass, $\Sigma m_{\nu}$, and the effective number of relativistic degrees of freedom, $N_{\mathrm{eff}}$, through cosmological observations \cite{Hu:1997mj},\cite{Feng:2019mym}, \cite{Yarahmadi}. These parameters play a critical role in shaping our understanding of the early universe's thermal history and the evolution of large-scale structures. Additionally, the cosmological implications of interactions between dark energy and dark matter have been extensively explored \cite{Amendola:1999er}, \cite{Kumar:2016zpg}, \cite{Luciano:2023}, \cite{Wang:2007}, \cite{Wang:2024}   ,\cite{Zimdahl:2001}  . However, our interest lies in the specific scenario where the role of neutrinos becomes particularly significant. The influence of neutrinos on the dynamics of the universe, especially in the context of their mass and interaction with other components, offers a rich field of study, potentially impacting the interpretation of cosmological data and the understanding of the universe's accelerating expansion.
The holographic dark energy (HDE) approach offers a compelling alternative scenario for the quantitative description of dark energy, grounded in the holographic principle \cite{Hooft} ,\cite{Susskind}. This principle posits that the number of degrees of freedom, which is directly related to entropy, scales with the area enclosing a system rather than its volume. By establishing a connection between the largest length scale in a quantum field theory and its ultraviolet (UV) cutoff \cite{Cohen:1998zx}, one can derive a vacuum energy of holographic origin. On cosmological scales, this vacuum energy manifests as dark energy \cite{Li:2004rb}, \cite{Wang:2016och}. The HDE model, therefore, provides a theoretical framework that links quantum field theory and gravitational physics, offering insights into the mysterious nature of dark energy and its role in the universe's accelerated expansion.

The application of the holographic principle in cosmology posits that entropy is directly proportional to the area of the event horizon, similar to the Bekenstein - Hawking entropy for black holes. However, recent work by Barrow has shown that quantum-gravitational effects can introduce complex, fractal-like features into the structure of black holes. This intricate structure leads to scenarios where a black hole can possess a finite volume but an infinite (or finite) area, thereby modifying the standard entropy expression for black holes \cite{Barrow:2004}. Barrow's modified entropy is given by:

\begin{equation}\label{eq1}
	S_B = \left(\frac{A}{A_0}\right)^\frac{2+\Delta}{2},
\end{equation}

where \( A \) represents the standard horizon area, \( A_0 \) is the Planck area, and \( \Delta \) is a new exponent representing the quantum-gravitational deformation. In particular cases, when \( \Delta = 0 \), the expression reduces to the standard holographic dark energy model, whereas \( \Delta = 1 \) corresponds to Barrow's holography entropy.

It is crucial to note that this quantum-gravitationally corrected entropy is distinct from the standard ``quantum-corrected'' entropy, which typically includes logarithmic corrections \cite{Kaul:2000kf}, \cite{Carlip:2000nv}. Although Barrow's entropy shares some resemblance with non-extensive Tsallis entropy \cite{Tsallis:1987eu}-\cite{Tsallis:2012js}, the underlying physical principles and foundational concepts are fundamentally different.

Recently, Saridakis \cite{Saridakis:2005}, by utilizing the usual holographic principle and applying Barrow entropy instead of the Bekenstein-Hawking entropy, reformulated the Barrow holographic dark energy (BHDE). Moreover, the case where \(\Delta = 0\) reduces BHDE to the standard HDE, although, in general, BHDE represents a new and broader framework with intriguing cosmological behavior. The standard HDE is defined by the inequality \(\rho_{B} L^{4} \leq S\), where \(L\) denotes the horizon length, and under the assumption that \(S \propto A \propto L^{2}\) \cite{Li:2004rb}, using the equation for Barrow entropy, Eq. (\ref{eq1}) can be modified as:

\begin{equation}\label{eq2}
	\rho_{B} = C L^{\Delta-2},
\end{equation}

where \(C\) is the holographic parameter with dimensions \([L]^{-2-\Delta}\) \cite{Saridakis:2005}. As expected, for \(\Delta=0\), the above expression reduces to the standard HDE expression \(\rho_{B} = 3 c^{2} M_{p}^{2} L^{-2}\), where \(C = 3 c^{2} M_{p}^{2}\) with \(c^{2}\) being the model parameter and \(M_{p}\) the Planck mass.

The Barrow holographic dark energy (BHDE) model has been applied in various cosmological contexts. Researchers have examined the BHDE model in flat Friedmann-Robertson-Walker (FRW) universes \cite{Pradhan:2021}, \cite{Shikha}, and in non-flat FRW universes \cite{Bhardwaj}, \cite{Dixit}. 

Moreover, a generalized entropy function has been introduced in \cite{Odintsov2022}, which can encompass all previously proposed entropies, including Barrow entropy. This generalized entropy function features four parameters and can represent Tsallis, Renyi, Barrow, Sharma-Mittal, Kaniadakis, and Loop Quantum Gravity entropies in appropriate limits. The study demonstrates that this generalized entropy framework effectively links early inflationary phases with the late dark energy era of the universe. It also adheres to the third law of thermodynamics, and the inflationary parameters \((n_{s}, r)\) and dark energy parameters \((\omega_{D}, q)\) are simultaneously satisfied within suitable ranges of the parameters \cite{Odintsov2022}.

Recent measurements of the Hubble constant, \(H_{0}\), derived from supernova observations, exhibit a substantial discrepancy compared to values obtained from cosmic microwave background (CMB) observations. Despite significant technological advancements and increased measurement accuracy, this discrepancy has grown to 5.3\(\sigma\) as precision has improved, contrary to expectations. This growing divergence suggests a potential need for new physics beyond the \(\Lambda\)CDM model, as it appears unrelated to measurement errors. There are two main methods for determining \(H_{0}\). The first, the direct measurement method, involves assessing the apparent recession velocities of galaxies and their distances. One significant challenge in this approach is obtaining accurate distance measurements. The second method uses CMB sound peaks with cosmological model constraints. While CMB observations provide valuable information about cosmological parameters, they require additional assumptions, such as a flat universe (\(\Omega_{K} = 0\)), to determine \(H_{0}\) from the CMB power spectrum observed by WMAP and Planck \cite{Pradhan:2021}, \cite{Shikha}, \cite{Bhardwaj}, \cite{Dixit}, \cite{Eleonora2021}.

Recent data from the SH0ES collaboration reports \(H_{0} = 74.03 \pm 1.42 \, \mathrm{km \, s^{-1} \, Mpc^{-1}}\). In contrast, Planck 2018 measurements give \(H_{0} = 67.4 \pm 0.5 \, \mathrm{km \, s^{-1} \, Mpc^{-1}}\) \cite{Planck2018}, illustrating the substantial tension between different observational methods.

Furthermore, another tension arises with the parameter \(S_8\), which quantifies the amplitude of matter density fluctuations in the universe. Defined as \(S_8 = \sigma_8 \sqrt{\frac{\Omega_m}{0.3}}\), where \(\sigma_8\) is the amplitude of fluctuations on scales of \(8h^{-1}\) Mpc, and $\Omega_{m}$ is the
matter density parameter at present time, \(S_8\) provides insights into the amplitude of density fluctuations and cosmological structure formation. Measurements from Planck CMB data yield \(S_8 = 0.832 \pm 0.013\) \cite{Planck2018}, yet this value shows a 2\(\sigma\) tension with measurements from galaxy clusters and weak lensing surveys. Specifically, KiDS-1000x\{2dFLenS+BOSS\} reports \(S_8 = 0.766^{+0.02}_{-0.014}\) \cite{51}, and DES-Y3 provides \(S_8 = 0.776 \pm 0.017\) \cite{50}. These discrepancies underscore ongoing challenges in aligning observational data with cosmological models.


\section{The model (Non interacting)}
Consider a spatially flat universe with HDE, matter, and radiation. The Friedmann equation
reads
 \begin{equation}\label{fried}
\begin{split}
3H^{2}M_{pl}^{2}=\rho.
\end{split}
\end{equation}
In the model, the total energy density and pressure are given by:

\begin{equation}
	\rho = \rho_{dm} + \rho_{B} + \rho_{b} + \rho_{\nu} + \rho_{r},
\end{equation}

\begin{equation}
	p = p_{m} + p_{B} + p_{\nu} + p_{r}.
\end{equation}

Here, \(\rho_{dm}\), \(\rho_{B}\), \(\rho_{b}\), \(\rho_{r}\), and \(\rho_{\nu}\) denote the energy densities of dark matter (DM), Barrow Holographic Dark Energy (BHDE), baryons, radiation, and neutrinos, respectively. The dark matter and baryons are considered as non-relativistic matter, while photons contribute to the radiation density. Neutrinos initially behave as relativistic particles, contributing to the radiation density in the early universe, but transition to a non-relativistic state at later times, acting as matter. Additionally, there may be an extra component of radiation often referred to as "dark radiation."

Additionally, for the matter sector we consider the standard conservation equation
\begin{equation}\label{eq4}
\dot{\rho}_i+3H\rho_i(1+\omega_i)=0.
\end{equation}
where $\omega_{1,2,3}=0$ for dark matter,baryon and neutrino masses, and $\omega_{4}=\frac{1}{3}$ for relativistic components.

 We introduce the density parameters
 \begin{eqnarray}\label{eq5}
 && \Omega_m\equiv\frac{\rho_m}{3H^2}, \Omega_{B}\equiv\frac{\rho_{B}}{3H^2} ,  \Omega_{\nu}\equiv\frac{\rho_{\nu}}{3H^2}
  ,  \Omega_r\equiv\frac{\rho_{r}}{3H^2}.
   \end{eqnarray}
From conservation law, we have,
\begin{equation}\label{eq5.1}
\dot{\rho}_{m}+3 H \rho_{m}=0,
\end{equation}
Where $\rho_{m}=\rho_{c}+\rho_{b}+\rho_{\nu}$,
\begin{equation}\label{eq5.2}
\dot{\rho}_{\nu}+3 H \rho_{\nu}=0,
\end{equation}
\begin{equation}\label{eq5.3}
\dot{\rho}_{r}+4 H \rho_{r}=0,
\end{equation}
Moreover, with derivation Eq. (2), If we take into
consideration the IR cut off $L$ as the Hubble horizon $H^{-1}$, then the energy density of BHDE
is obtained as
 \begin{equation}\label{eq5.4}
\rho_{B}=CH^{2-\Delta}.
\end{equation}
Hence,
 \begin{equation}\label{eq5.4}
\frac{\dot{\rho}_{B}}{3H^{3}}=(2-\Delta)\frac{\dot{H}}{H^{2}}\Omega_{B},
\end{equation}
  Hence, the equations of the autonomous dynamical system can be derived as,

 \begin{align}\label{eq6}
&\frac{d\Omega_m}{dN}=-3\Omega_m-2\frac{\dot{H}}{H^{2}}\Omega_m,\\ \nonumber
&\frac{d\Omega_{B}}{dN}=-\Delta\frac{\dot{H}}{H^{2}}\Omega_{B},\\ \nonumber
&\frac{d\Omega_{\nu}}{dN}=-3\Omega_{\nu}-2\frac{\dot{H}}{H^{2}}\Omega_{\nu},\\ \nonumber
&\frac{d\Omega_r}{dN}=-4\Omega_r-2\frac{\dot{H}}{H^{2}}\Omega_r, \\ \nonumber
\end{align}
where, $N=\ln a$. Since $3H^{2}=\rho$, then $6H\dot{H}=\dot{\rho}=$ In term of the new dynamical variable, for IR cut of Hubble horizon, we can obtain
\begin{equation}\label{hh}
\begin{split}
\epsilon=\frac{\dot{H}}{H^{2}}=\frac{-3\Omega_m-4\Omega_r}{2+(\Delta-2)\Omega_{B}}.
\end{split}
\end{equation}
Also in terms of the new variables the Friedmann equation (\ref{fried}) puts a constraint on the new variables as
\begin{equation}\label{eq7}
\begin{split}
\Omega_m+\Omega_{B}+\Omega_r=1.
\end{split}
\end{equation}
The parameter $\epsilon$ for $\Delta=0$ would be
 \begin{equation}\label{eqq}
\begin{split}
 \epsilon=\frac{-3\Omega_m-4\Omega_r}{2-2\Omega_{B}}.
\end{split}
\end{equation}
Since, the deceleration parameter can be obtained as
 \begin{equation}q=-\Big(1+\epsilon\Big).\end{equation}
 Hence the expansion of the universe is accelerating if $\epsilon>-1$.
This condition can be satisfied if
\begin{equation}
\frac{3\Omega_m+4\Omega_r}{2-2\Omega_{B}}<1,
\end{equation}
 which can be simplified as
\begin{equation}\label{re}
\frac{3}{2}\Omega_m+2\Omega_r+\Omega_{B}<1.
\end{equation}
By rewriting the left hand side of the equation (\ref{re}) in terms of $\Omega_{1}$ and $\Omega_{2}$ as
\begin{align}
&\Omega_{1}=\Omega_m+\Omega_r+\Omega_{B},\\
&\Omega_{2}=\frac{1}{2}\Omega_m+\Omega_r.
\end{align}
The condition (\ref{re}) can be rewritten as
\begin{align}\label{re2}
\Omega_{1}+\Omega_{2}<1.
\end{align}
Note that from equation (\ref{eq7}), where \(\Omega_{1} = 1\), it is evident that \(\Omega_{2} > 0\). Consequently, the condition given by equations (\ref{re}) or (\ref{re2}) cannot be satisfied. Thus, we conclude that the standard holographic dark energy model with the Hubble horizon as the infrared (IR) cutoff is insufficient to support the accelerating expansion of the universe.

This issue has been highlighted in previous studies \cite{Hsu, Mauricio, Li}. Specifically, using the current Hubble horizon as the IR cutoff in the Friedmann equation \(\rho = 3M_{pl}^{2} H^{2}\) causes the dark energy to behave more like non-relativistic matter rather than a fluid with negative pressure. As a result, it obstructs the possibility of an accelerating universe.

In this scenario, we find that both the matter density \(\rho_{m} \sim H^{2}\) and the dark energy density \(\rho_{DE} \sim H^{2}\) scale similarly. This tracking behavior indicates that dark matter and holographic dark energy evolve in tandem, resulting in a dark energy component that approximates pressureless matter rather than exhibiting the negative pressure required for cosmic acceleration.

However for $\Delta\neq0$, the accelerating condition is as
\begin{equation}
\begin{split}
\epsilon=\frac{3\Omega_m+4\Omega_r}{2+(\Delta-2)\Omega_{B}}<1,
\end{split}
\end{equation}
which can be simplified as
\begin{equation}\label{re4}
\frac{3}{2}\Omega_m+2\Omega_r+\Omega_{B}-\frac{\Delta}{2}\Omega_{B}<1,
\end{equation}
which can be separated as
\begin{align}
&\Omega_{1}=\Omega_m+\Omega_r+\Omega_{B},\\
&\Omega_{2}=-\frac{\Delta}{2}\Omega_{B}+\frac{1}{2}\Omega_m+\Omega_r,
\end{align}
since $\Omega_{1}=1$, the accelerating condition is $\Omega_{2}<0$ which can be simplified as

\begin{equation}\label{re4}
\Delta>\frac{\Omega_m+2\Omega_r}{\Omega_{B}}.
\end{equation}
However, the best value of the parameter $\Delta$ could be obtained using observational data. Hence in the rest of the paper, we try to find the best value for parametters of the model.

In order to put
constraints on the parameters of the model we must note that the model has four independent variables $(\Omega_{BH},\Omega_{\nu},\Omega_{r},\Omega_{dm})$ as well as two free parameters of the model $(\omega,\Delta)$. Hence in order to solve the equation numerically the four initial conditions $(\Omega_{BH}(0),\Omega_{\nu}(0),\Omega_{r}(0),\Omega_{dm}(0))$ and value of the parameters must be known.  The other parameters are expressed in terms of the main parameters and can be constrained indirectly.  For example
 $\sum m_{\nu}$ can be related to the main parameters $(h,\Omega_{\nu})$ as
\begin{eqnarray}
\Omega_{\nu}=\frac{\sum m_{\nu}}{94 h^{2}eV},
\end{eqnarray}
where  $h$ is the reduced Hubble constant (the Hubble constant $H_{0} = 100h$ km$ s^{-1}$ Mp$c^{-1}$. Hence if the parameters $(h,\Omega_{\nu})$ are constrained then the parameter $\sum m_{\nu}$ is constrained automatically.

The relativistic energy density in the early universe include the contributions
from photons and neutrinos, and possibly other extra relativistic
degrees of freedom, called dark radiation. The effective
number of relativistic species, including neutrinos and any other
dark radiation, is defined by a parameter, $N_{eff}$ , for which the standard
value is 3.046 corresponding to the case with three-generation
neutrinos and no extra dark radiation \cite{Mangano}. We
parameterize the relativistic degrees of freedom using
the effective number of neutrino species, $N_{eff}$. This quantity can be written in
terms of the matter density,
$\Omega_{m}h^{2}$, and the redshift of
matter-radiation equality $z_{eq}$ as \cite{Komatsu:2011}
\begin{equation}
\begin{split}
N_{eff}=3.04+7.44\Big(\frac{\Omega_{m}h^{2}}{0.1308}\frac{3139}{1+z_{eq}}-1\Big).
\end{split}
\end{equation}

In following, We derive the Hubble parameter \( H(z) \) for a universe containing matter, radiation, neutrinos, and Barrow Holographic Dark Energy (BHDE). The starting point for this derivation is the Friedmann equation, which governs the dynamics of a spatially flat universe.

The energy densities of different components as functions of redshift \( z \) can be expressed compactly as follows. For matter (dark matter + baryons), radiation, and neutrinos, the energy densities evolve as \(\rho_m(z) = \rho_{m0}(1+z)^3\), \(\rho_r(z) = \rho_{r0}(1+z)^4\), and \(\rho_\nu(z) = \rho_{\nu0}(1+z)^4\), respectively, where \(\rho_{m0}\), \(\rho_{r0}\), and \(\rho_{\nu0}\) represent their present-day energy densities. Assuming neutrinos remain relativistic at early times, they share the same redshift dependence as radiation. The energy density of Barrow Holographic Dark Energy (BHDE) is given by \(\rho_B(z) = C H(z)^{2-\Delta}\), where \(C\) is a constant and \(\Delta\) is the Barrow exponent parameter.

The Friedmann equation can now be rewritten by substituting the energy densities:
\begin{equation}
	\begin{split}
		&H^2(z) = \frac{1}{3 M_{\text{pl}}^2}[\rho_{m0}(1+z)^3 + \rho_{r0}(1+z)^4 +\\& \rho_{\nu0}(1+z)^4 + C H(z)^{2-\Delta}].
	\end{split}
\end{equation}

Substituting these definitions into the Friedmann equation, we obtain:
\begin{equation}
	E^2(z) = \Omega_{m0}(1+z)^3 + \Omega_{r0}(1+z)^4 + \Omega_{\nu0}(1+z)^4 + \Omega_{B0} E(z)^{-\Delta},
\end{equation}
where \( E(z) = \frac{H(z)}{H_0} \) is the dimensionless Hubble parameter.

This equation shows that the dimensionless Hubble parameter \( E(z) \) depends on the redshift \( z \) and the parameters of the different components of the universe. The equation is generally solved numerically since \( E(z) \) appears on both sides of the equation. Once \( E(z) \) is determined, the Hubble parameter \( H(z) \) is given by:
\begin{equation}
	H(z) = H_0 E(z).
\end{equation}

For small redshifts \( z \), where \( 1+z \approx 1 \), the radiation contribution becomes negligible, and the Hubble parameter can be approximated as:
\begin{equation}
	H(z) \approx H_0 \sqrt{\Omega_{m0}(1+z)^3 + \Omega_{B0} E(z)^{-\Delta}}.
\end{equation}
This approximation is useful for low-redshift analyses, but the full numerical solution is required for precise cosmological modeling across a broader range of redshifts.

\section{Barrow Interacting Holographic dark energy)}
In the previous section, we have investigated the Barrow holographic dark energy (BHDE) model and put constraints on the model parameter $\Delta$. While the BHDE model is generally effective in describing the evolution of the Universe, a notable issue arises concerning the value of the parameter $\Delta$. According to the theoretical framework of BHDE, the parameter $\Delta$ should lie within the interval $0 < \Delta < 1$. However, our statistical analysis has yielded a best-fit value of $\Delta < 0.43$ (We use cut off form 0)at 68\% confidence level (CL), which clearly falls outside the expected range.
Another solution is that the model improved by interacting term. Hence
in this section we study the holographic dark energy with interacting term $Q=3\beta H\rho_{B}$,
From conservation law, we have,
\begin{equation}\label{eq5.1}
\dot{\rho}_{m}+3 H \rho_{m}=Q,
\end{equation}
\begin{equation}\label{eq5.1}
\dot{\rho}_{B}+3 H \rho_{B}=-Q,
\end{equation}
\begin{equation}\label{eq5.2}
\dot{\rho}_{\nu}+3 H \rho_{\nu}=0,
\end{equation}
\begin{equation}\label{eq5.3}
\dot{\rho}_{r}+4 H \rho_{r}=0,
\end{equation}
Hence,
 \begin{equation}\label{eq5.4}
\frac{\dot{\rho}_{B}}{3H^{3}}=(2-\Delta)\frac{\dot{H}}{H^{2}}\Omega_{B},
\end{equation}
Hence, under the condition $\Delta-3\beta>\frac{\Omega_m+2\Omega_r}{\Omega_{B}}$ the model can describe the late time acceleration of the Universe.
 The equations of the autonomous dynamical system can be derived as,
 \begin{align}\label{eq6}
&\frac{d\Omega_m}{dN}=-3\Omega_m-3\beta\Omega_{B}-2\frac{\dot{H}}{H^{2}}\Omega_m,\\ \nonumber
&\frac{d\Omega_{B}}{dN}=-\Delta\frac{\dot{H}}{H^{2}}\Omega_{B},\\ \nonumber
&\frac{d\Omega_{\nu}}{dN}=-3\Omega_{\nu}-2\frac{\dot{H}}{H^{2}}\Omega_{\nu},\\ \nonumber
&\frac{d\Omega_r}{dN}=-4\Omega_r-2\frac{\dot{H}}{H^{2}}\Omega_r ,\\ \nonumber
\end{align}
where, $N=\ln a$. Since $3H^{2}=\rho$, then $6H\dot{H}=\dot{\rho}=$ In term of the new dynamical variable, for IR cut of Hubble horizon, we can obtain
\begin{equation}
\begin{split}
\epsilon=\frac{\dot{H}}{H^{2}}=\frac{-3\Omega_m-3\beta\Omega_{B}-4\Omega_r}{2+(\Delta-2)\Omega_{B}}.
\end{split}
\end{equation}

\section{Derivation of the Hubble Parameter in the Interacting BHDE Model}

In this section, we derive the Hubble parameter \( H(z) \) for a cosmological model where the Barrow Holographic Dark Energy (BHDE) interacts with dark matter through a term \( Q = 3\beta H \rho_{B} \). This interaction modifies the conservation equations and consequently affects the evolution of the Hubble parameter.

The Friedmann equation, which relates the Hubble parameter to the energy densities of the universe, is given by:
\begin{equation}
	H^2 = \frac{1}{3 M_{\text{pl}}^2}\left(\rho_m + \rho_B + \rho_\nu + \rho_r\right), \label{eq:friedmann}
\end{equation}
which can be rewritten in terms of the density parameters as:
\begin{equation}
	\begin{split}
		H^2(z) = H_0^2 & \left[\Omega_{m0}(1+z)^3 + \Omega_{B0} E(z)^{-\Delta} \right. \\
		& \left. + \Omega_{\nu0}(1+z)^4 + \Omega_{r0}(1+z)^4\right], \label{eq:hubble_param}
	\end{split}
\end{equation}

where \( E(z) = \frac{H(z)}{H_0} \) is the dimensionless Hubble parameter, and \( \Omega_{m0}, \Omega_{B0}, \Omega_{\nu0}, \Omega_{r0} \) are the present-day density parameters for matter, BHDE, neutrinos, and radiation, respectively.

Finally, by substituting the expression for \(\epsilon\) from Equation 40 into the expression for \( H(z) \), we obtain the full equation for \( E(z) \) as:
\begin{equation}
	\begin{split}
		E(z) = & \sqrt{\Omega_{m0}(1+z)^3 + \Omega_{B0} E(z)^{-\Delta} +} \\
		& \sqrt{\Omega_{\nu0}(1+z)^4 + \Omega_{r0}(1+z)^4}. 
	\end{split}
	\label{eq:hubble_solution}
\end{equation}

This equation typically requires numerical methods to solve for \( E(z) \), given that \( E(z) \) appears on both sides of the equation. The final expression for the Hubble parameter \( H(z) \) thus depends on the redshift \( z \), the coupling parameter \( \beta \), and the Barrow exponent \( \Delta \), reflecting the effects of the interaction between BHDE and dark matter.


\section{Observational Datasets and the statistical Methodology}
To evaluate the success of the model under study, we perform a series of Markov-chain Monte Carlo (MCMC) runs, using the public code { MontePython-v3}\footnote{\url{https://github.com/brinckmann/montepython_public}}\cite{54,55}, which we interface with our modified version of { CLASS}~\cite{56,57}. To test the convergence of the MCMC chains, we use the Gelman-Rubin \cite{58} criterion $|R -1|\!\lesssim\!0.01$. To post-process the chains and plot figures we use {\sf GetDist} \cite{59}. To fit the perturbed Hubble parameter model using Monte Carlo methods with data from CMB, (CC), (BAO), Pantheon supernovae, and Lensing one can use the MontePython package, which is specifically designed for cosmological parameter estimation. Begin by defining the Hubble parameter function within your model, which includes perturbations and various cosmological parameters. Set up the MontePython environment by creating a configuration file that specifies the cosmological model, the parameters to be fitted, and the observational data to be used. MontePython will then handle the integration with different observational datasets, allowing you to incorporate CMB measurements, BAO data, supernovae observations, and cosmic chronometer constraints into the fitting process. The configuration file will also define the prior distributions for each parameter based on theoretical expectations or observational limits.

Once the configuration is complete, use MontePython to perform the Markov Chain Monte Carlo (MCMC) sampling. The package employs sophisticated sampling techniques to explore the parameter space, providing posterior distributions for each parameter. Analyze the MCMC results by examining the chains' convergence and calculating statistical metrics such as mean values, credible intervals, and likelihoods. MontePython's output will include a range of diagnostic plots and statistical summaries, which are crucial for evaluating the fit of your model to the data. By comparing the model predictions with the observational data, you can assess the accuracy and robustness of your Hubble parameter model.
All observational data where used in this paper are:\\
$\bullet$ Pantheon catalog:
We used updated the Pantheon + Analysis catalog consisting of 896 SNe Ia covering the redshift range $0.12 < z < 2.3$ \cite{Scolnic2022}. Pantheon data is powerful data for investigating about anisotropy in late universe \cite{Yar3} 

MontePython uses Pantheon supernovae data by comparing the theoretical distance moduli with observed moduli. The likelihood function $\mathcal{L}_{\text{SN}}$ for supernovae data is expressed as:
\begin{equation}
	\mathcal{L}_{\text{SN}} \propto \exp\left(-\frac{1}{2} \sum_{j} \left[ \frac{\mu_{j}^{\text{model}} - \mu_{j}^{\text{data}}}{\sigma_{\mu}(j)} \right]^2 \right),
\end{equation}
where $\mu_{j}^{\text{model}}$ is the predicted distance modulus for the $j$-th supernova, $\mu_{j}^{\text{data}}$ is the observed distance modulus, and $\sigma_{\mu}(j)$ is the uncertainty. MontePython calculates the likelihood for each parameter set by comparing the model's predicted distance moduli with the observed values.\\

$\bullet$ {CMB data}:
We used the latest large-scale cosmic microwave background (CMB) temperature and
polarization angular power spectra from the final release of Planck 2018 plikTTTEEE+lowl+lowE
\cite{Planck2018}. \\
MontePython utilizes CMB data to constrain cosmological models by comparing theoretical predictions of the CMB power spectra with observed power spectra. The likelihood function for CMB data, $\mathcal{L}_{\text{CMB}}$, is based on the deviation between the observed and theoretical power spectra. It is given by:
\begin{equation}
	\mathcal{L}_{\text{CMB}} \propto \exp\left(-\frac{1}{2} \left[ \textbf{d}_{\text{obs}} - \textbf{d}_{\text{theory}} \right]^T \textbf{C}^{-1} \left[ \textbf{d}_{\text{obs}} - \textbf{d}_{\text{theory}} \right] \right),
\end{equation}
where $\textbf{d}_{\text{obs}}$ and $\textbf{d}_{\text{theory}}$ represent the observed and theoretical CMB power spectra, respectively, and $\textbf{C}$ is the covariance matrix accounting for measurement uncertainties and correlations. MontePython incorporates precomputed CMB likelihood functions, typically provided by analysis pipelines such as those from the Planck mission.\\

$\bullet$ {Lensing}: we consider the 2018 CMB lensing reconstruction power spectrum data,
obtained with a CMB trispectrum analysis in \cite{Aghanim1}.
MontePython integrates lensing data by evaluating the theoretical predictions of the lensing power spectrum against observed data. The likelihood function for lensing data, $\mathcal{L}_{\text{lensing}}$, is expressed as:
\begin{align}
	\mathcal{L}_{\text{lensing}} &\propto \exp\left(-\frac{1}{2} \left[ \textbf{d}_{\text{lensing,obs}} - \textbf{d}_{\text{lensing,theory}} \right]^T \right. \nonumber \\
	&\quad \times \left. \textbf{C}_{\text{lensing}}^{-1} \left[ \textbf{d}_{\text{lensing,obs}} - \textbf{d}_{\text{lensing,theory}} \right] \right),
\end{align}

where $\textbf{d}_{\text{lensing,obs}}$ and $\textbf{d}_{\text{lensing,theory}}$ are the observed and theoretical lensing power spectra, respectively, and $\textbf{C}_{\text{lensing}}$ is the covariance matrix incorporating measurement uncertainties and correlations. This data is crucial for understanding the distribution of dark matter and the large-scale structure of the universe.\\

$\bullet$ {BAO data}:
We also used various measurements of BAO data (\cite{Carter2018,Gil-Marin2020,Bautista2021,DES2022,Neveux2020,Hou2021,Bourboux2020}) are in \cite{Ratra1}
\begin{table}
	\centering
	\caption{12 BAO data.}\label{tab:bao}
	\scriptsize
	\begin{tabular}{lccc}
		\hline
		$z$ & Measurement{a} & Value \\
		\hline
		\hline
		$0.122$ & $D_V\left(r_{s,{\rm fid}}/r_s\right)$ & $539\pm17$ \\
		$0.38$ & $D_M/r_s$ & 10.23406 \\
		$0.38$ & $D_H/r_s$ & 24.98058 \\
		$0.51$ & $D_M/r_s$ & 13.36595 \\
		$0.51$ & $D_H/r_s$ & 22.31656 \\
		$0.698$ & $D_M/r_s$ & 17.858 \\
		$0.698$ & $D_H/r_s$ & 19.326 \\
		$0.835$ & $D_M/r_s$ & $18.92\pm0.51$ \\
		$1.48$ & $D_M/r_s$ & 30.6876 \\
		$1.48$ & $D_H/r_s$ & 13.2609 \\
		$2.334$ & $D_M/r_s$ & 37.5 \\
		$2.334$ & $D_H/r_s$ & 8.99 \\
		\hline
	\end{tabular}
\end{table}
where $D_V$, $r_s$, $r_{s, {\rm fid}}$, $D_M$, $D_H$, and $D_A$ have units of Mpc.
MontePython incorporates BAO data by fitting theoretical predictions of the BAO scale with observed measurements. The likelihood function $\mathcal{L}_{\text{BAO}}$ is given by:
\begin{equation}
	\mathcal{L}_{\text{BAO}} \propto \exp\left(-\frac{1}{2} \frac{\left(D_V^{\text{model}} - D_V^{\text{data}}\right)^2}{\sigma_{D_V}^2}\right),
\end{equation}
where $D_V^{\text{model}}$ is the model’s predicted BAO scale, $D_V^{\text{data}}$ is the observed scale, and $\sigma_{D_V}$ is the uncertainty in the measurement. In MontePython, users input observed BAO scales and their uncertainties, and the software computes the likelihood based on the fit between theoretical predictions and observations.

 The 12 BAO measurements listed in Table \ref{tab:bao} cover the redshift range $0.122 \leq z \leq 2.334$. The quantities listed in Table 1 are described as follows:

 $D_V(z)$: Spherically averaged BAO distance, $D_V(z)=[czH(z)^{-1}D^2_M(z)]^{1/3}$, where $c$ is the speed of light and the angular diameter distance $D_A(z) = D_M(z)/(1+z)$ with $D_M(z)$ defined in the following\\
$D_H(z)$: Hubble distance, $D_H(z)=c/H(z)$\\
$r_s$: Sound horizon at the drag epoch, $r_{s, {\rm fid}}=147.5$ Mpc.\\
 $D_M(z)$: Transverse comoving distance,
\begin{equation}
	\label{eq:DM}
	D_M(z) = 	D_C(z) \ \ \ \  \text{if}\ \Omega_{k0} = 0,\\
\end{equation}
where the comoving distance
\begin{equation}
	\label{eq:gz}
	D_C(z) = c\int^z_0 \frac{dz'}{H(z')}.
\end{equation}
\\
$\bullet$ {CC data}: The cosmic chronometer (CC) data covering the redshift $0.07 < z < 1.965$.(Table 2)\\

\begin{table}[t!]
	\scriptsize
	\centering
	\caption{32 $H(z)$ (CC) data.}\label{tab:hz}
	\begin{tabular}{lcc}
		\hline
		$z$ & $H(z)$ & $\sigma$\\
		\hline
		0.07 & $69.0$ & 19.6\\
		0.09 & $69.0$ & 12.0\\
		0.12 & $68.6$ & 26.2\\
		0.17 & $83.0$ & 8.0\\
		0.2 & $72.9$ & 29.6\\
		0.27 & $77.0$ & 14.0\\
		0.28 & $88.8$ & 36.6\\
		0.4 & $95.0$ & 17.0\\
		0.47 & $89.0$ & 50.0\\
		0.48 & $97.0$ & 62.0\\
		0.75 & $98.8$ & 33.6\\
		0.88 & $90.0$ & 40.0\\
		0.9 & $117.0$ & 23.0\\
		1.3 & $168.0$ & 17.0\\
		1.43 & $177.0$ & 18.0\\
		1.53 & $140.0$ & 14.0\\
		1.75 & $202.0$ & 40.0\\
		0.1791 & 74.91 & 4.00\\
		0.1993 & 74.96 & 5.00\\
		0.3519 & 82.78 & 14\\
		0.3802 & 83.0 &  13.5\\
		0.4004 & 76.97 &  10.2\\
		0.4247 & 87.08 &  11.2\\
		0.4497 & 92.78 &  12.9\\
		0.4783 & 80.91 &  9\\
		0.5929 & 103.8 & 13\\
		0.6797 & 91.6 & 8\\
		0.7812 & 104.5 & 12\\
		0.8754 & 125.1 & 17\\
		1.037 & 153.7 & 20\\
		1.363 & 160.0 & 33.6\\
		1.965 & 186.5 & 50.4\\
		\hline
	\end{tabular}
\end{table}
MontePython uses Cosmic Chronometers data, which provides measurements of the expansion rate $H(z)$ at various redshifts, to constrain model parameters. The likelihood function $\mathcal{L}_{\text{CC}}$ is expressed as:
\begin{equation}
	\mathcal{L}_{\text{CC}} \propto \exp\left(-\frac{1}{2} \sum_{i} \left[ \frac{H(z_i)_{\text{model}} - H(z_i)_{\text{data}}}{\sigma_{H}(z_i)} \right]^2 \right),
\end{equation}
where $H(z_i)_{\text{model}}$ denotes the predicted expansion rate at redshift $z_i$, $H(z_i)_{\text{data}}$ is the observed expansion rate, and $\sigma_{H}(z_i)$ is the uncertainty. MontePython integrates these measurements by specifying them in the configuration file and computing the likelihood based on model predictions.

In MontePython, the likelihood functions for all data types are combined to form the total likelihood function:
\begin{equation}
	\mathcal{L}(\text{data} | \theta) = \mathcal{L}_{\text{CMB}} \times \mathcal{L}_{\text{CC}} \times \mathcal{L}_{\text{BAO}} \times \mathcal{L}_{\text{SN}} \times \mathcal{L}_{\text{lensing}},
\end{equation}
where $\theta$ represents the cosmological parameters.

MontePython performs Markov Chain Monte Carlo (MCMC) sampling to explore the parameter space. The posterior probability distribution is given by:
\begin{equation}
	\mathcal{P}(\theta | \text{data}) \propto \mathcal{L}(\text{data} | \theta) \cdot \text{Prior}(\theta),
\end{equation}
where $\text{Prior}(\theta)$ is the prior distribution for the parameters. MontePython iterates over parameter values, proposing new sets of parameters, and accepting or rejecting them based on their likelihood. This process generates a chain of parameter samples, from which posterior distributions and constraints on the parameters are derived.

Table III demonstrate flat priors for the cosmological parameters used in the analysis of non-interacting and interacting BHDE models.
\begin{table}[h!]
	\centering
	\caption{Flat priors for the cosmological parameters used in the analysis of non-interacting and interacting BHDE models.}
	\begin{tabular}{c c}
		\hline 
		Parameter                    & Prior\\
		\hline
		$\Omega_{b} h^2$             & $[0.005, 0.1]$\\
		$\Omega_{c} h^2$             & $[0, 0.15]$\\
		$\tau$                       & $[0.01, 0.8]$\\
		$n_s$                        & $[0.8, 1.2]$\\
		$\log[10^{10}A_{s}]$         & $[1.6, 3.9]$\\
		$100\theta_{MC}$             & $[0.5, 10]$\\ 
		${\Omega _\nu}$              & $[0, 0.01]$\\
		\hline 
	\end{tabular}
	\label{tab:priors}
\end{table}
\section{Results and discussion}
To evaluate the discrepancies between the Hubble constant measurements, we utilize the tension equation, which quantifies the degree of disagreement between different datasets and the Planck 2018 and R22 results. This is expressed in terms of standard deviations (\(\sigma\)) from the Planck 2018 and R22 results.

The table provides a detailed comparison of the Hubble constant (\(H_0\)) across various datasets, presenting the values of \(H_0\) along with their associated uncertainties and tensions relative to the Planck 2018 and R22 measurements. The values are expressed in \(\text{km/s/Mpc}\) For non- interacting model, with the uncertainties representing the standard deviation of each measurement.

	The tension between the different \(H_0\) measurements and the values from Planck 2018 and R22 is calculated using the following formula:
\begin{equation}
		\sigma_{\text{tension}} = \frac{|H_0^{\text{dataset}} - H_0^{\text{reference}}|}{\sqrt{\sigma_{H_0, \text{dataset}}^2 + \sigma_{H_0, \text{reference}}^2}}.
\end{equation}

Based on this relation, We obtained the Hubble tension froem Planck 2018 and R22 and demonstrated them in table IV.

\begin{table}
	\scriptsize
	\centering
	\begin{tabular}{|c|c|c|c|}
		\hline
		\textbf{Dataset Combination} & \textbf{Tension with Planck 2018} & \textbf{Tension with R22} & \textbf{\(H_0\) Value (km/s/Mpc)} \\ 
		&  &  & \\ 
		\hline
		CMB + Lensing & $0.62\sigma$ & $1.27\sigma$ & $69.5 \pm 2.9$ \\ 
		\hline
		CMB + BAO + Lensing & $1.24\sigma$ & $1.19\sigma$ & $70.11 \pm 1.7$  \\ 
		\hline
		CMB + Lensing + CC & $1.17\sigma$ & $1.23\sigma$ & $70.5 \pm 1.93$ \\ 
		\hline
		CMB + Lensing + Pantheon & $1.24\sigma$ & $1.12\sigma$ &$ 70.9 \pm 1.93$ \\ 
		\hline
		CMB + All  & $1.52\sigma$ & $1.45\sigma$ & $70.21 \pm 1.8 $ \\ 
		\hline
	\end{tabular}
	\caption{Tension between different \(H_0\) measurements with the values from Planck 2018 and R22 for the non-interacting model.}
	\label{tab:tension_results}
\end{table}

Figure 1 presents a comparative analysis of the Hubble constant (\( H_{0} \)) estimates derived from various combinations of observational datasets within the framework of the Non-Interacting Barrow Holographic Dark Energy model. For detailed numerical values and corresponding uncertainties of \( H_{0} \) across these dataset combinations, refer to Table IV. This figure effectively illustrates the variation in \( H_{0} \) as influenced by different data constraints, providing insight into the model's performance in addressing the Hubble tension. 
\begin{figure*}
	\includegraphics[scale=.5]{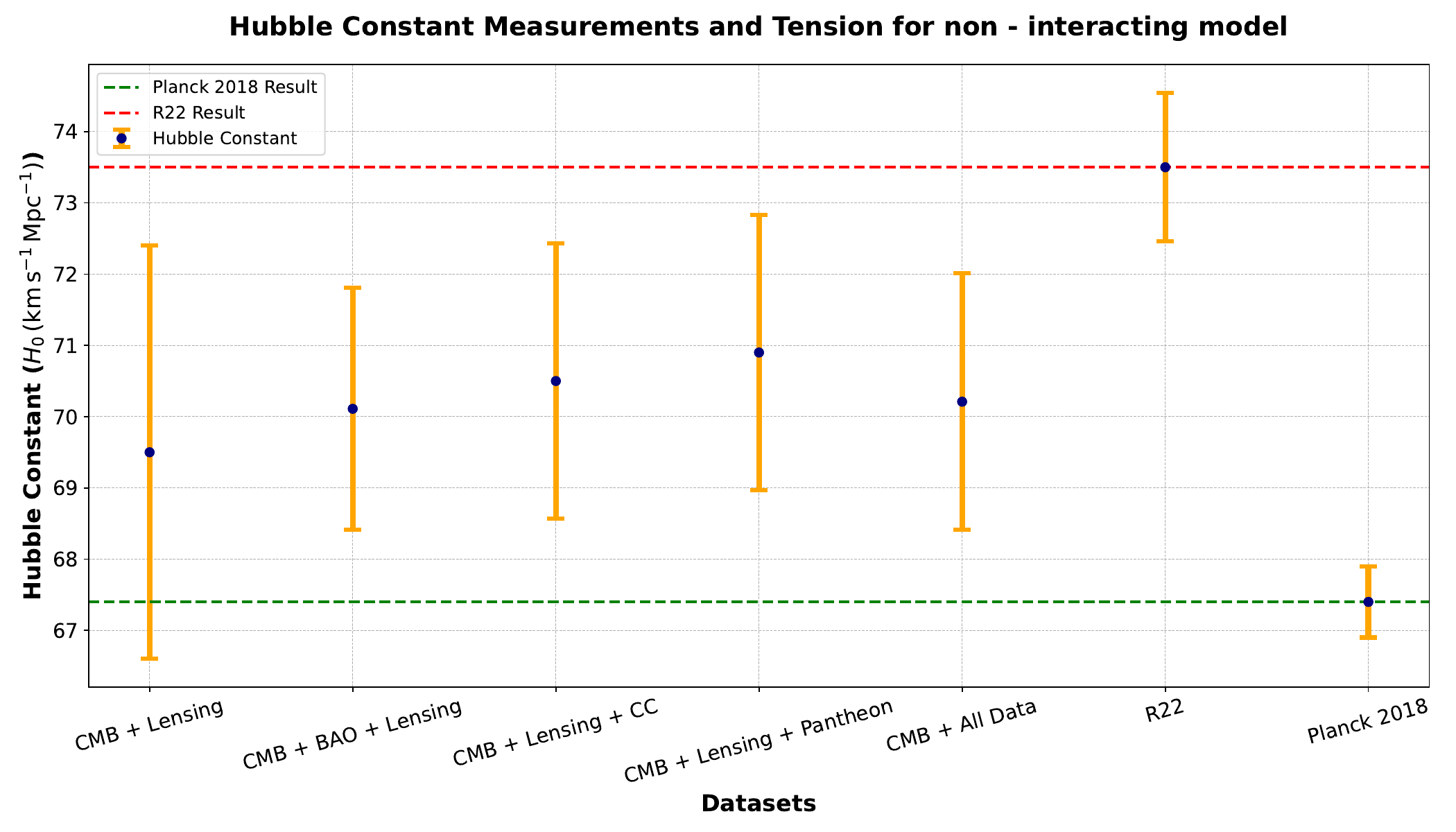}\hspace{0.1 cm}\\
	Fig. 1: The comparison of  $H_{0}$ results for different combination of datasets for  Non-Interacting Barrow Holographic dark energy \\
\end{figure*}

The \( S_8 \) parameter, which combines the matter density \( \Omega_m \) and the amplitude of matter fluctuations \( \sigma_8 \), is crucial for understanding the large-scale structure of the Universe. In our analysis, we find the \( S_8 \) value for the CMB + All datasets to be \( S_8 = 0.791 \pm 0.032 \).

For comparison, the Planck CMB measurement yields \( S_8 = 0.832 \pm 0.013 \)~\cite{Planck2018}. This indicates a noticeable tension of approximately \( 1.28\sigma \) with the Planck results.

Additionally, when comparing our findings to measurements from KiDS-1000x\{2dFLenS+BOSS\}, which reports \( S_8 = 0.766^{+0.02}_{-0.014} \)~\cite{50}, and DES-Y3, which provides \( S_8 = 0.776 \pm 0.017 \)~\cite{51}, the tension with these datasets is also significant. For the CMB + All results, the tension with KiDS-1000 is approximately \( 1.35\sigma \) and with DES-Y3 is about \( 1.67\sigma \).

These findings underscore the challenges in reconciling cosmic microwave background data with large-scale structure observations, highlighting the level of discrepancy between our non-interacting model and various observational datasets.
 All the above results are in table V.
\begin{table}
	\centering
	\begin{tabular}{|c|c|c|c|c|c|}
		\hline
		\textbf{Dataset} & \textbf{Value} & \textbf{Planck Tension} & \textbf{KiDS-1000 Tension} & \textbf{DES-Y3 Tension} \\
		\hline
		CMB + Lensing & \( S_8 = 0.798 \pm 0.067 \) & \( 0.50\sigma \) & \( 0.45\sigma \) & \( 0.62\sigma \) \\
		\hline
		CMB + Lensing + BAO & \( S_8 = 0.794 \pm 0.047 \) & \( 0.77\sigma \) & \( 0.85\sigma \) & \( 1.06\sigma \) \\
		\hline
		CMB + Lensing + CC & \( S_8 = 0.790 \pm 0.031 \) & \( 1.35\sigma \) & \( 1.42\sigma \) & \( 1.77\sigma \) \\
		\hline
		CMB + Lensing + Pantheon & \( S_8 = 0.793 \pm 0.049 \) & \( 0.68\sigma \) & \( 0.75\sigma \) & \( 0.94\sigma \) \\
		\hline
		CMB + All & \( S_8 = 0.791 \pm 0.032 \) & \( 1.28\sigma \) & \( 1.35\sigma \) & \( 1.67\sigma \) \\
		\hline
	\end{tabular}
	\caption{Summary of \( S_8 \) tensions with Planck 2018, KiDS, and DES for the non-interacting model.}
	\label{tab:S8_results}
\end{table}
Figure 2, demonstrate the comparison $S_{8}$ for different combination of datasets for  Non-Interacting Barrow Holographic dark energy
\begin{figure*}
	\includegraphics[scale=.6]{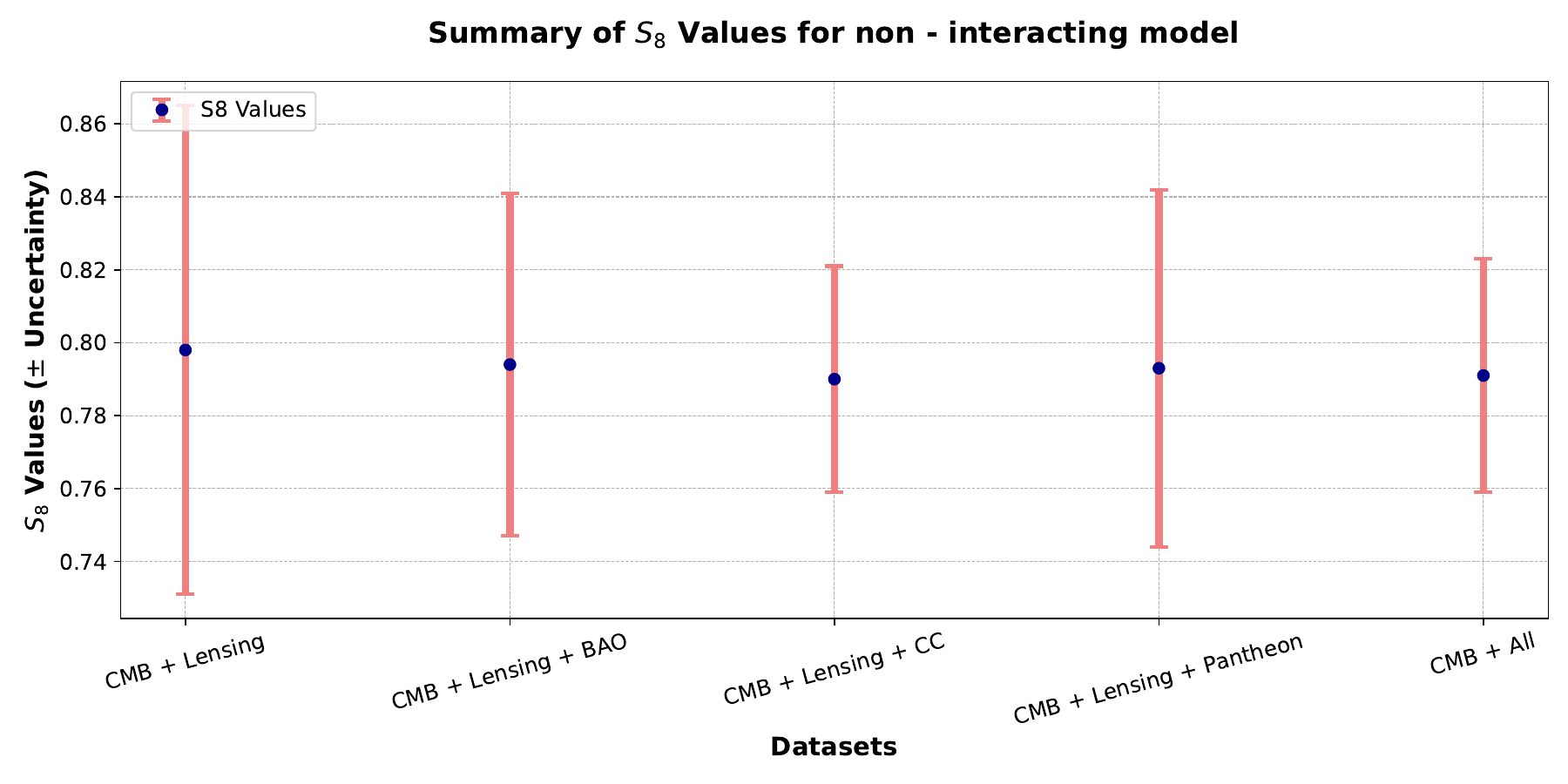}\hspace{0.1 cm}\\
	Fig. 2: The comparison $S_{8}$ for different combination of datasets for  Non-Interacting Barrow Holographic dark energy \\
\end{figure*}

The results obtained from the non-interacting and interacting BHDE models, as presented in Table VI, are visually depicted in Figures 3-7. The table provides a comprehensive summary of how different observational datasets constrain the main and derived cosmological parameters in the non-interacting BHDE model. The results demonstrate the model's ability to provide consistent and robust constraints, with the combined dataset (CMB+All) offering the most stringent bounds. The reported upper limits on the neutrino mass sum and the effective number of relativistic species are in line with current cosmological expectations, suggesting that the non-interacting BHDE model is a viable framework for describing the universe's evolution. These figures provide a clear graphical representation of the cosmological parameters and their respective constraints, allowing for a more intuitive comparison of the model predictions with observational data. Furthermore, Table VI specifically highlights the cosmological parameters derived from the non-interacting BHDE model. The corresponding results are illustrated in Figure 8, offering a detailed visualization of how these parameters align with current observations. These figures and tables together underscore the effectiveness of both models in addressing the $H_0$ and $S_8$ tensions, while also providing a comprehensive overview of the parameter space explored in this study. The "CMB+ALL" results are especially important, as they offer the most stringent constraints and are likely to be the most reliable for interpreting the model's implications for cosmic history.

\begin{table*}[t]   
	\caption{Cosmological Parameter Results and Observational Constraints at $68\%$ for the Non-Interacting Model. The parameter $H_0$ is in units of $km/s/Mpc$, and $\sum m_{\nu}$ reported at the $95\%$ CL is in units of eV.}
	\centering
	\resizebox{1\textwidth}{!}{
		\begin{tabular}{|l|c|c|c|c|c|}
			\hline
			Parameter & CMB+Pantheon+Lensing & CMB+CC+Lensing & CMB+BAO+Lensing & CMB+All & CMB+Lensing \\
			\hline
			${\Omega_b h^2}$ & $0.02225 \pm 0.00022$ & $0.02227 \pm 0.00023$ & $0.02235 \pm 0.00019$ & $0.02222 \pm 0.00022$ & $0.02237 \pm 0.0002$ \\
			\hline
			${\Omega_c h^2}$ & $0.1197 \pm 0.0035$ & $0.1193 \pm 0.0036$ & $0.1192 \pm 0.0034$ & $0.1190 \pm 0.0025$ & $0.1194 \pm 0.0034$ \\
			\hline
			${100\theta_{MC}}$ & $1.04088 \pm 0.00054$ & $1.04087 \pm 0.00055$ & $1.04113 \pm 0.00046$ & $1.04099 \pm 0.00051$ & $1.04114 \pm 0.00045$ \\
			\hline
			${\tau}$ & $0.0548 \pm 0.0076$ & $0.0545 \pm 0.0076$ & $0.0553 \pm 0.0080$ & $0.0553^{+0.0054}_{-0.0078}$ & $0.05527 \pm 0.0081$ \\
			\hline
			${\mathrm{ln}(10^{10} A_s)}$ & $3.045 \pm 0.017$ & $3.046 \pm 0.016$ & $3.044 \pm 0.018$ & $3.046^{+0.017}_{-0.018}$ & $3.046 \pm 0.018$ \\
			\hline
			${n_s}$ & $0.9685 \pm 0.0085$ & $0.9675 \pm 0.0085$ & $0.9647 \pm 0.0075$ & $0.9655 \pm 0.0075$ & $0.9645 \pm 0.0076$ \\
			\hline
			$\Delta$ & $<0.59$ & $<0.56$ & $<0.45$ & $<0.43$ & $<0.46$ \\
			\hline
			$\Omega_m$ & $0.296^{+0.007}_{-0.0062}$ & $0.295^{+0.008}_{-0.009}$ & $0.296^{+0.016}_{-0.013}$ & $0.296^{+0.006}_{-0.005}$ & $0.322^{+0.012}_{-0.017}$ \\
			\hline
			$\sum m_{\nu}$ (eV) & $<0.181$ & $<0.2$ & $<0.124$ & $<0.122$ & $<0.276$ \\
			\hline
			$N_{eff}$ & $3.01^{+0.23}_{-0.22}$ & $2.99^{+0.22}_{-0.26}$ & $2.98^{+0.29}_{-0.28}$ & $2.99^{+0.19}_{-0.2}$ & $2.95^{+0.37}_{-0.32}$ \\
			\hline
		\end{tabular}
	}
	\label{table:merged}
\end{table*}

\begin{figure}
	\includegraphics[scale=.35]{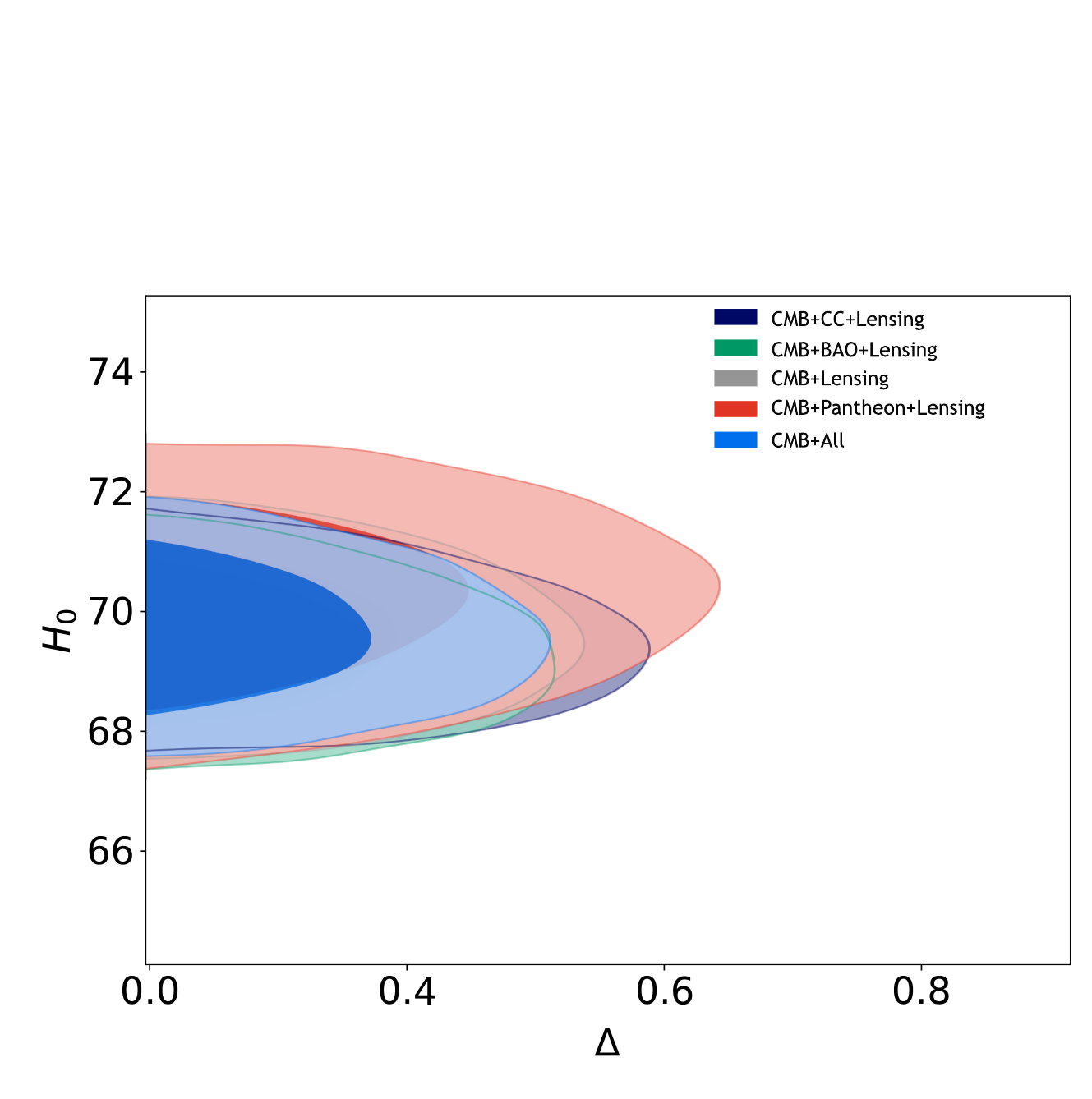}\hspace{0.1 cm}\\
	Fig. 3: The comparison of  $\Delta$ parameter according to $H_{0}$ for different combination of datasets for  Non-Interacting Barrow Holographic dark energy \\
\end{figure}

\begin{figure}
	\includegraphics[scale=.35]{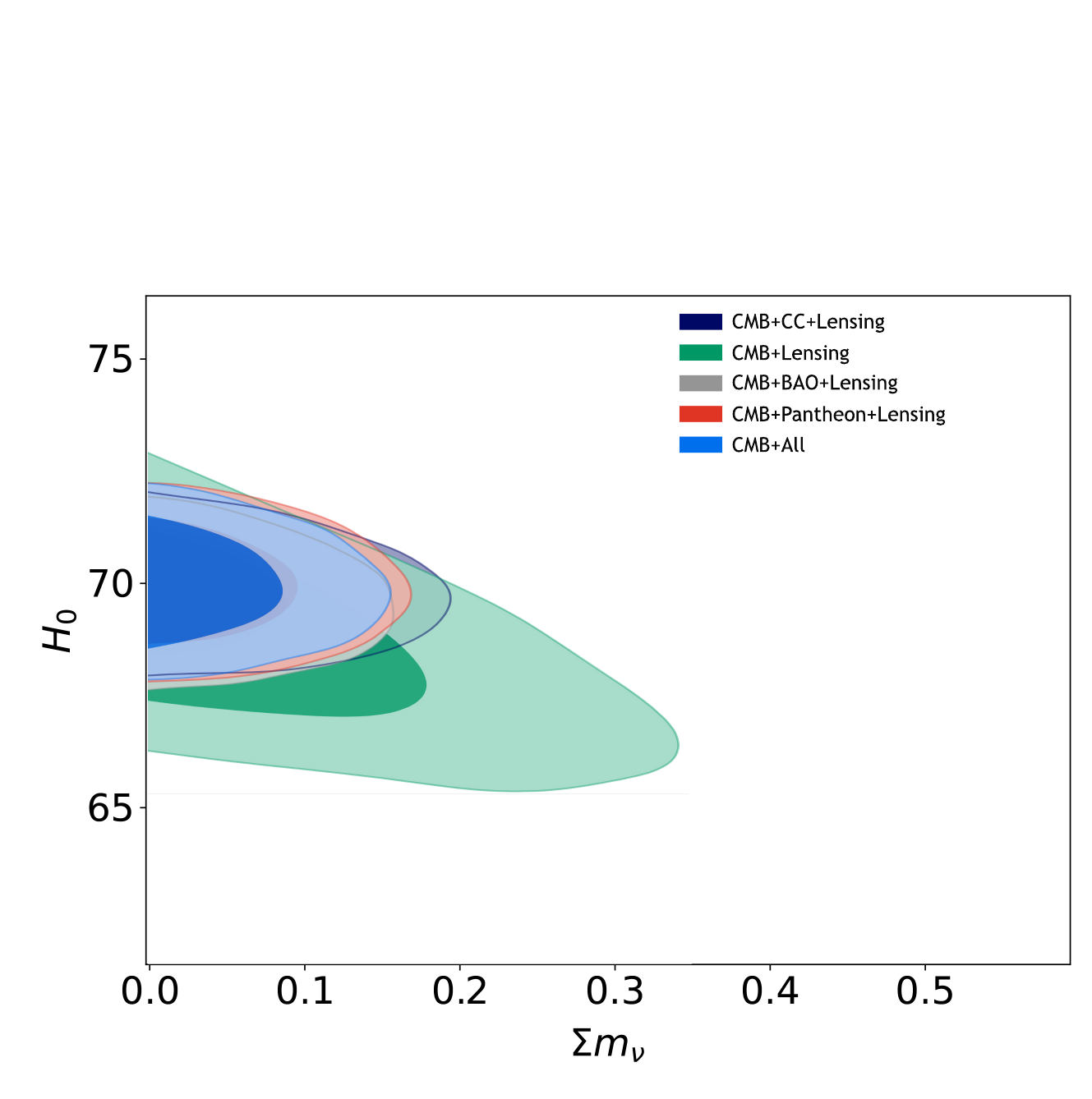}\hspace{0.1 cm}\\
	Fig. 4: Constraints at the (95$\% $CL.) two-dimensional contours for $\sum m_{\nu}$ for different combination of datasets in Non-Interacting Barrow Holographic dark energy \\
\end{figure}

\begin{figure}
	\includegraphics[scale=.35]{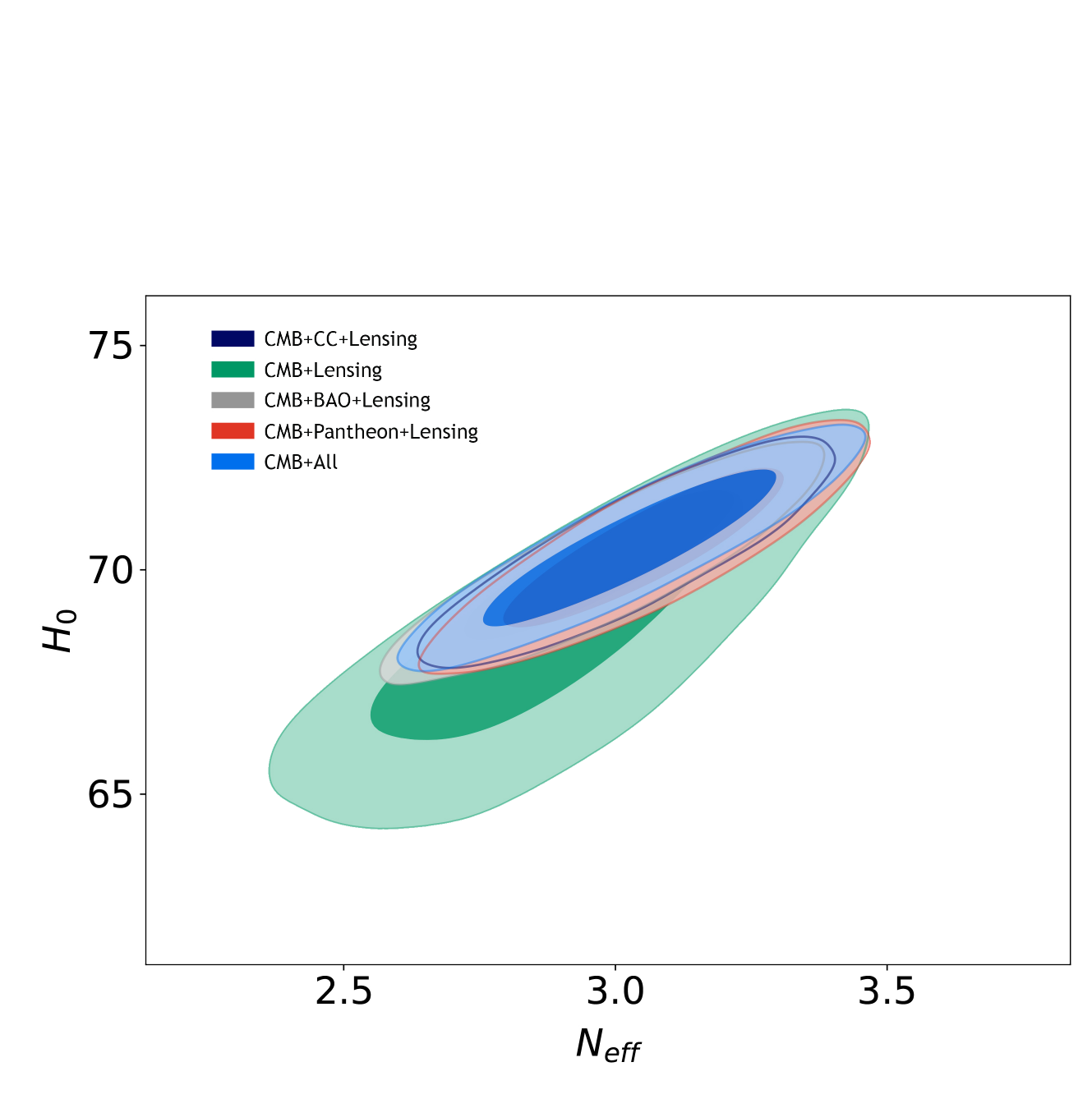}\hspace{0.1 cm}\\
	Fig. 5: Constraints at the (68$\% $CL.) two-dimensional contours for $N_{eff}$ for different combination of datasets in Non-Interacting Barrow Holographic dark energy \\
\end{figure}

\begin{figure}
	\includegraphics[scale=.35]{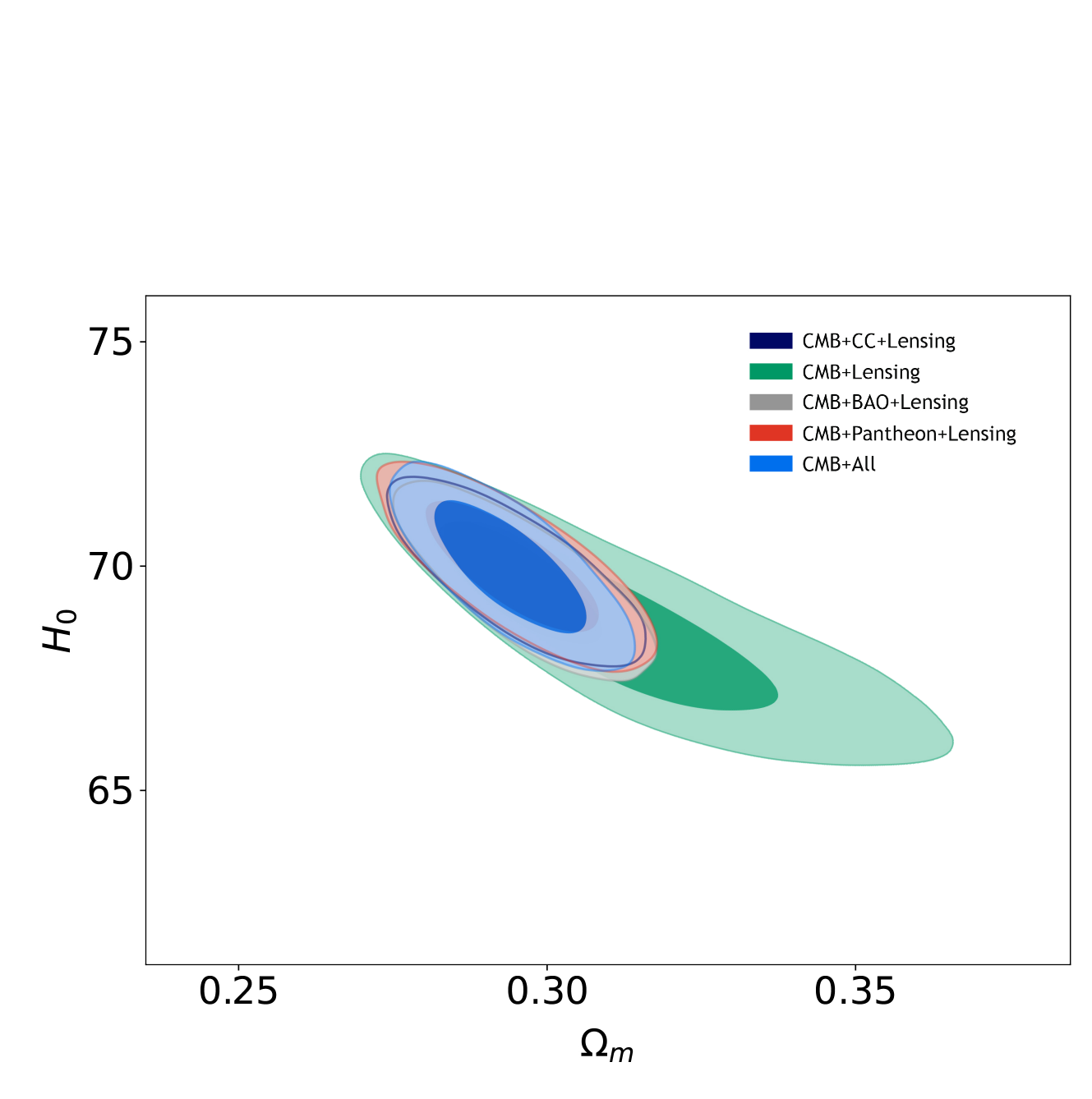}\hspace{0.1 cm}\\
	Fig. 6: Comparison  results of the  $\Omega_{m}$  according to $H_{0}$ for different combination dataset for Non-Interacting Barrow Holographic dark energy \\
\end{figure}

\begin{figure}
	\includegraphics[scale=.35]{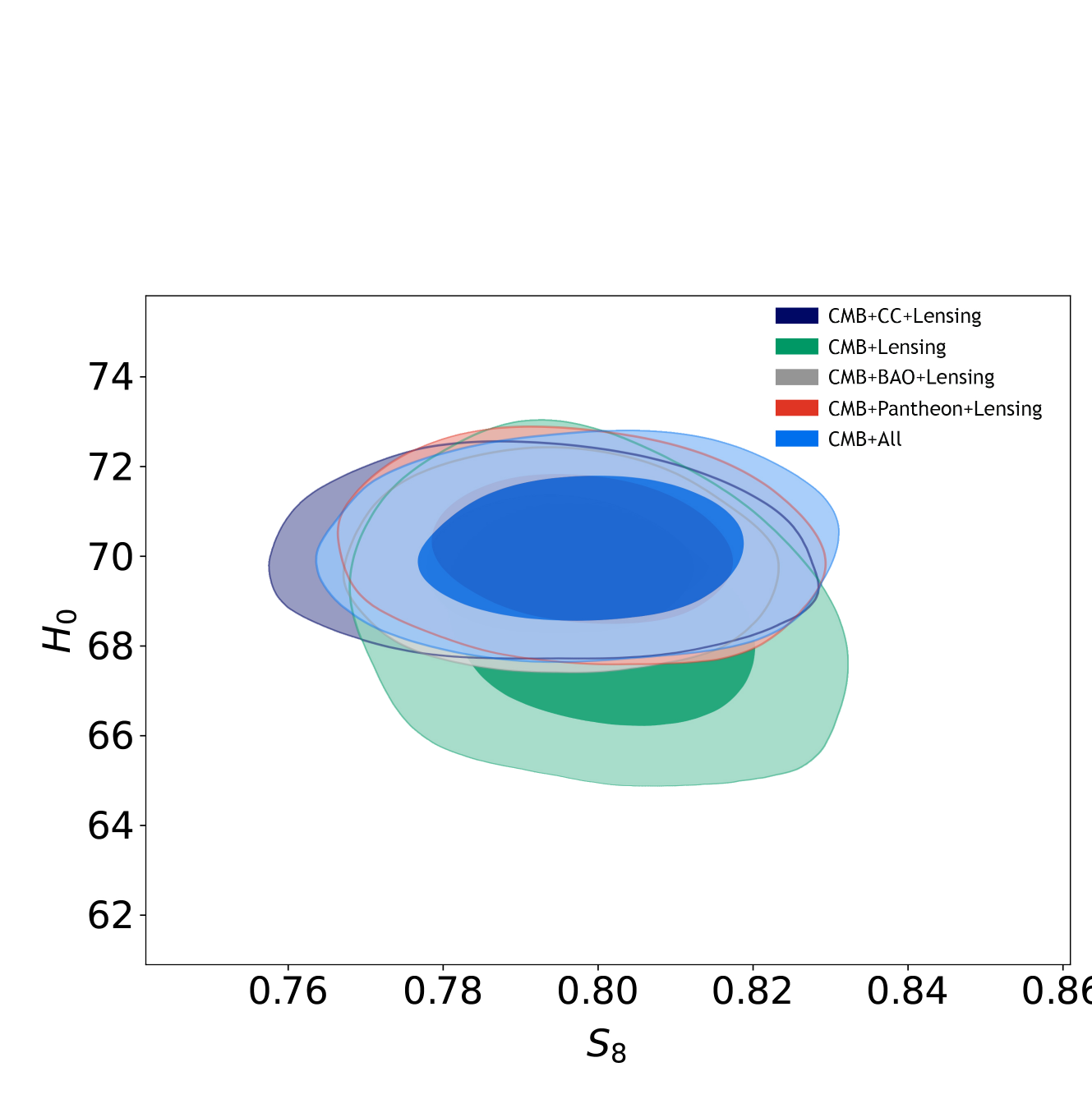}\hspace{0.1 cm}\\
	Fig. 7: Comparison  results of the  $S_{8}$ according to $H_{0}$ for different combination datasets for Non-Interacting Barrow Holographic dark energy \\
\end{figure}

\begin{figure*}
	\includegraphics[scale=.65]{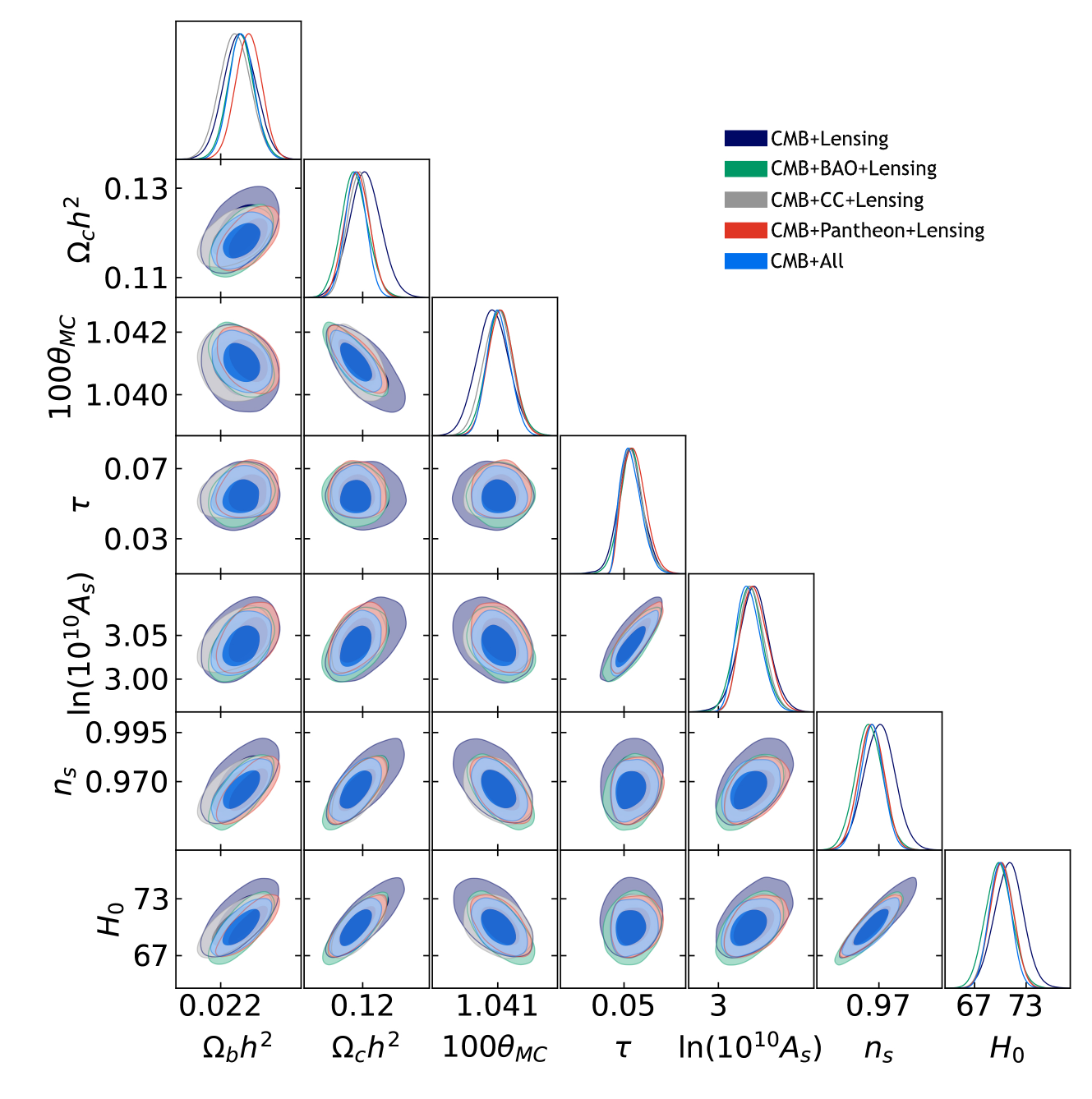}\hspace{0.1 cm}\\
	Fig. 8: Comparison  results of the  $S_{8} $ , $\Omega_b h^2   $, $\Omega_c h^2   $, $n_s  $, ${\rm{ln}}(10^{10} A_s)$, $100\theta_{MC} $, $H_{0}(km/s/Mpc
	)$ for different combination dataset for Non-Interacting Barrow Holographic dark energy \\
\end{figure*}
By using  the first and the second scenario, we can  put constraints on the following cosmological parameters: the Baryon energy density ${\Omega _b}{h^2}$, the cold dark matter energy density $\Omega_{c}h^{2}$, the neutrino density ${\Omega _\nu}$,  the ratio of the sound horizon at decoupling to the angular diameter distance to last scattering $\theta_{MC}$, the optical depth to reionization $\tau$, the amplitude and the spectral index of the primordial scalar perturbations $A_{s}$ and $n_{s}$. The results  obtained for two scenarios are in Tables VI and X.\\

\noindent The \( H_0 \) tension for the interacting model is evaluated using several dataset combinations, including CMB + Lensing, CMB + BAO + Lensing, CMB + Lensing + CC, CMB + Lensing + Pantheon, and CMB + All Data (BAO, CC, Pantheon, Lensing). For the CMB + Lensing dataset, the \( H_0 \) value \( (69.83 \pm 3.12 \, \text{km/s/Mpc}) \) shows a moderate tension of \( 0.96\sigma \) with R22 \( (73.04 \pm 1.04 \, \text{km/s/Mpc}) \), and a lower tension of \( 0.75\sigma \) with Planck 2018 \( (67.4 \pm 0.5 \, \text{km/s/Mpc}) \). When BAO data is included (CMB + BAO + Lensing), the tension with R22 increases slightly to \( 1.04\sigma \), and with Planck 2018 to \( 1.15\sigma \). Adding CC data (CMB + Lensing + CC) yields tensions of \( 0.99\sigma \) with R22 and \( 1.14\sigma \) with Planck. Incorporating Pantheon data (CMB + Lensing + Pantheon) results in a tension of \( 0.89\sigma \) with R22 and \( 1.18\sigma \) with Planck. The combination of all datasets (CMB + BAO + CC + Pantheon + Lensing) gives the highest tension with R22 \( (1.13\sigma) \) and Planck 2018 \( (1.43\sigma) \), suggesting that the full dataset combination indicates a modest discrepancy with both the R22 and Planck 2018 results. \noindent The results of these tensions for the \( H_0 \) values obtained from different dataset combinations are summarized in Table VII. These results are very close to \cite{Y10}

\begin{table}
	\centering
	\begin{tabular}{|c|c|c|c|c|}
		\hline
		\textbf{Dataset} & \textbf{Value} & \textbf{Tension with R22} & \textbf{Tension with Planck 2018} \\
		\hline
		CMB + Lensing & \( H_0 = 69.83 \pm 3.12 \) km/s/Mpc & \( 0.96\sigma \) & \( 0.75\sigma \) \\
		\hline
		CMB + BAO + Lensing & \( H_0 = 70.38 \pm 2.32 \) km/s/Mpc & \( 1.04\sigma \) & \( 1.15\sigma \) \\
		\hline
		CMB + Lensing + CC & \( H_0 = 70.27 \pm 2.38 \) km/s/Mpc & \( 0.99\sigma \) & \( 1.14\sigma \) \\
		\hline
		CMB + Lensing + Pantheon & \( H_0 = 70.58 \pm 2.54 \) km/s/Mpc & \( 0.89\sigma \) & \( 1.18\sigma \) \\
		\hline
		CMB + All  & \( H_0 = 70.47 \pm 2.01 \) km/s/Mpc & \( 1.13\sigma \) & \( 1.43\sigma \) \\
		\hline
	\end{tabular}
	\caption{Summary of \( H_0 \) tensions with Planck 2018 and R22 for the interacting model.}
	\label{tab:H0_results}
\end{table}
Figure 9 showcases a comparative evaluation of the Hubble constant (\( H_{0} \)) obtained from different combinations of observational datasets within the framework of the Interacting Barrow Holographic Dark Energy model. This comparison highlights the impact of dataset combinations on the inferred \( H_{0} \) values, providing a comprehensive understanding of how interactions within the dark energy sector influence the model's capability to reconcile the Hubble tension.

\begin{figure*}
	\includegraphics[scale=.5]{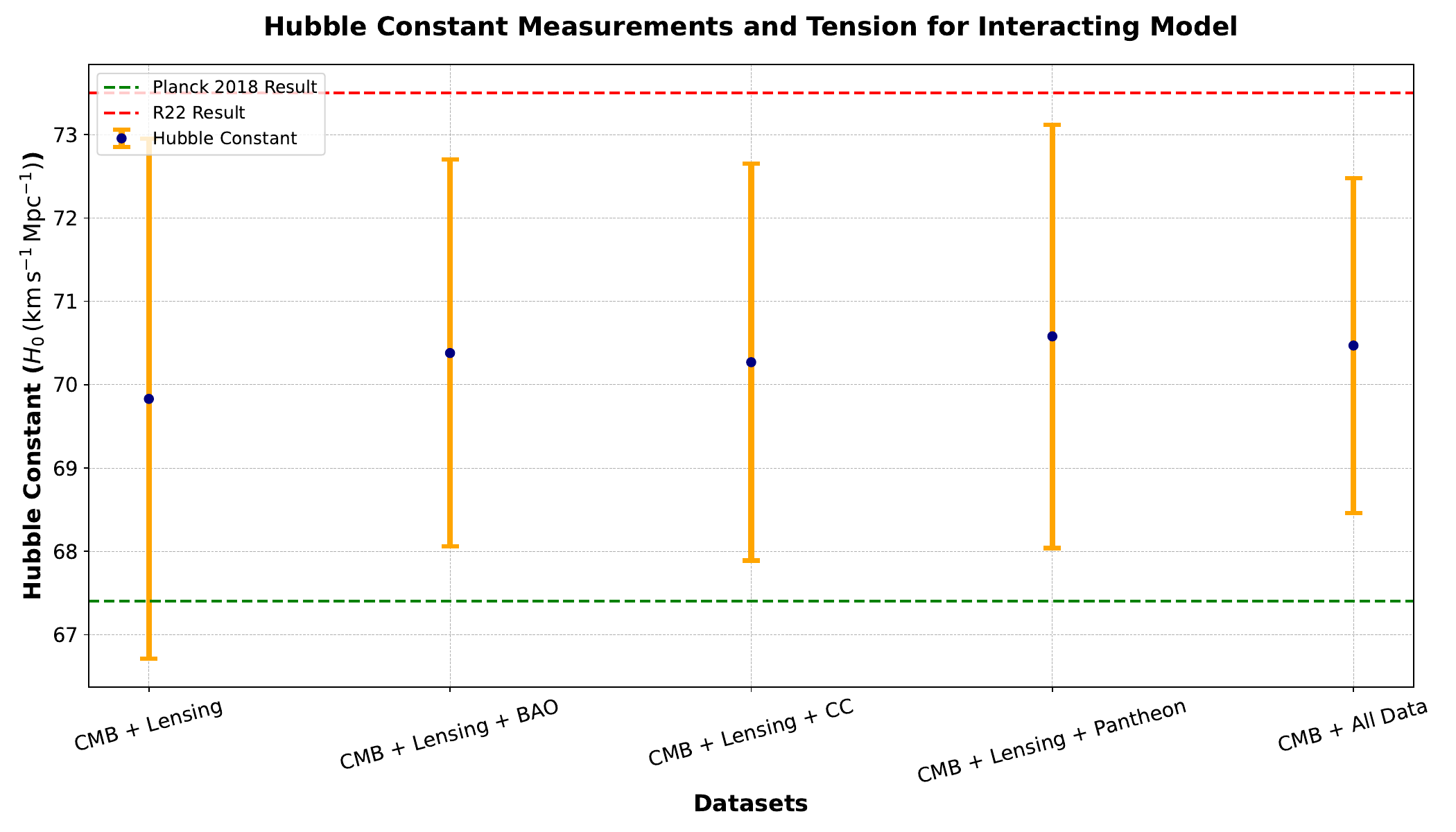}\hspace{0.1 cm}\\
	Fig. 9: Comparison  results of  $H_{0}$ for different combination datasets for Interacting Barrow Holographic dark energy \\
\end{figure*}

The values of \( S_8 \) derived from various combinations of datasets in our analysis are summarized in Table VIII.

\begin{table}
	\scriptsize
	\centering
	\begin{tabular}{|c|c|c|c|c|}
		\hline
		\textbf{Dataset} & \textbf{Value} & \textbf{Tension with Planck} & \textbf{Tension with KiDS-1000} & \textbf{Tension with DES-Y3} \\
		\hline
		CMB + Lensing & \( S_8 = 0.785 \pm 0.067 \) & \( 0.70\sigma \) & \( 0.86\sigma \) & \( 0.48\sigma \) \\
		\hline
		CMB + Lensing + BAO & \( S_8 = 0.783 \pm 0.047 \) & \( 0.99\sigma \) & \( 0.97\sigma \) & \( 0.59\sigma \) \\
		\hline
		CMB + Lensing + CC & \( S_8 = 0.784 \pm 0.048 \) & \( 0.89\sigma \) & \( 0.89\sigma \) & \( 0.51\sigma \) \\
		\hline
		CMB + Lensing + Pantheon & \( S_8 = 0.785 \pm 0.059 \) & \( 0.68\sigma \) & \( 0.90\sigma \) & \( 0.52\sigma \) \\
		\hline
		CMB + All & \( S_8 = 0.783 \pm 0.047 \) & \( 0.99\sigma \) & \( 0.97\sigma \) & \( 0.59\sigma \) \\
		\hline
	\end{tabular}
	\caption{Summary of \( S_8 \) tensions with Planck 2018, KiDS-1000, and DES-Y3 for the interacting model.}
	\label{tab:S8_results}
\end{table}

The results presented in Table IX are visually depicted in Figures 10 through 15, providing a comprehensive illustration of the derived cosmological parameters across various dataset combinations for the Interacting Barrow Holographic Dark Energy (BHDE) model. Additionally, the outcomes summarized in Table IX are illustrated in Figure 16. This figure offers a comparative analysis of key cosmological parameters, including $S_{8}$, $\Omega_b h^2$, $\Omega_c h^2$, $n_s$, ${\rm{ln}}(10^{10} A_s)$, $100\theta_{MC}$, and $H_{0}$ (in km/s/Mpc), across different dataset combinations. These comparisons highlight the variations in parameter constraints depending on the specific dataset utilized, thereby providing deeper insights into the effects of data combination on the parameter estimation within the interacting BHDE framework.

\begin{figure}
	\includegraphics[scale=.35]{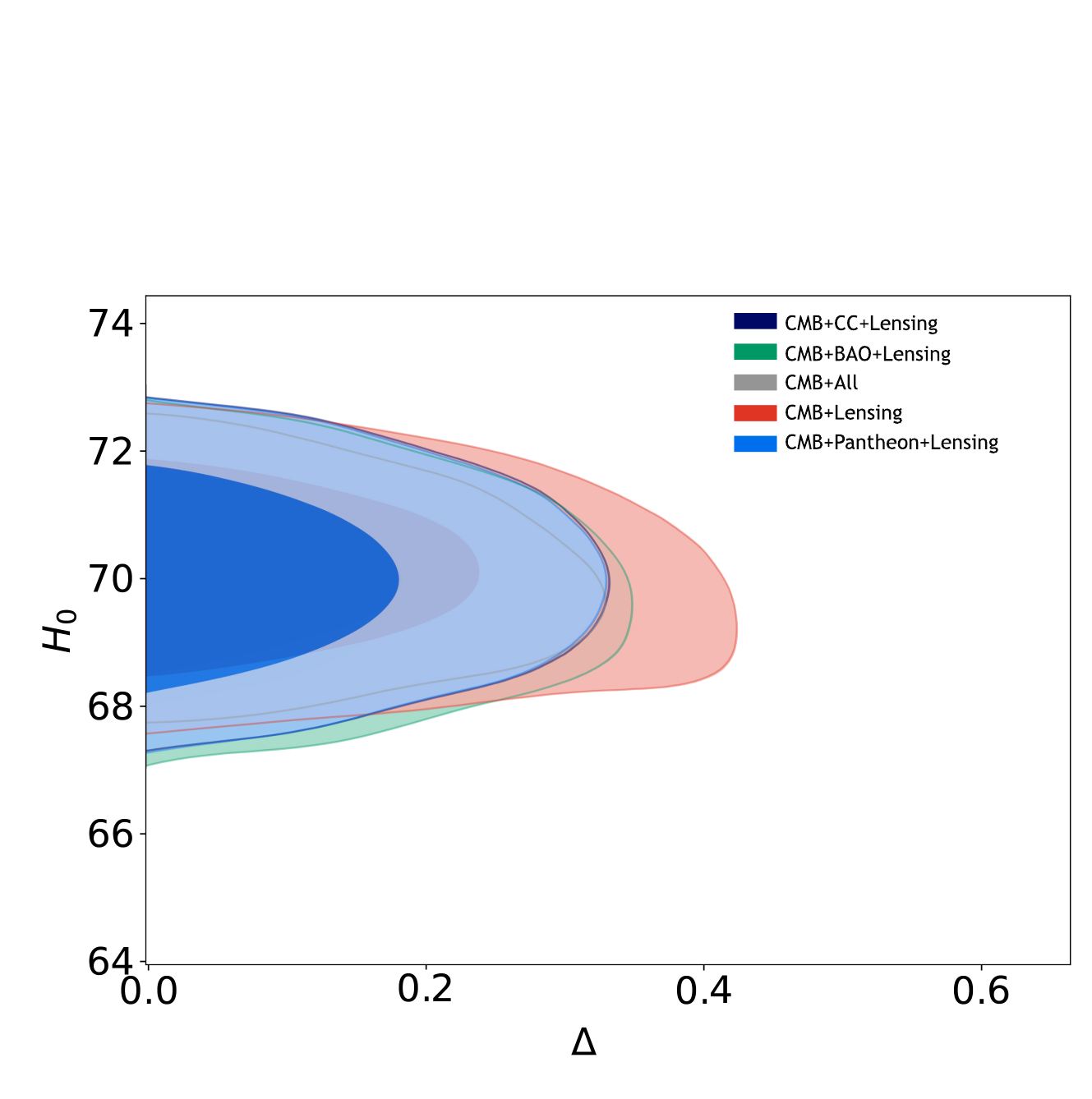}\hspace{0.1 cm}\\
	Fig. 10: The comparison of  $\Delta$ parameter according to $H_{0}$ for different combination of datasets for  Interacting Barrow Holographic dark energy \\
\end{figure}
\begin{figure}
	\includegraphics[scale=.35]{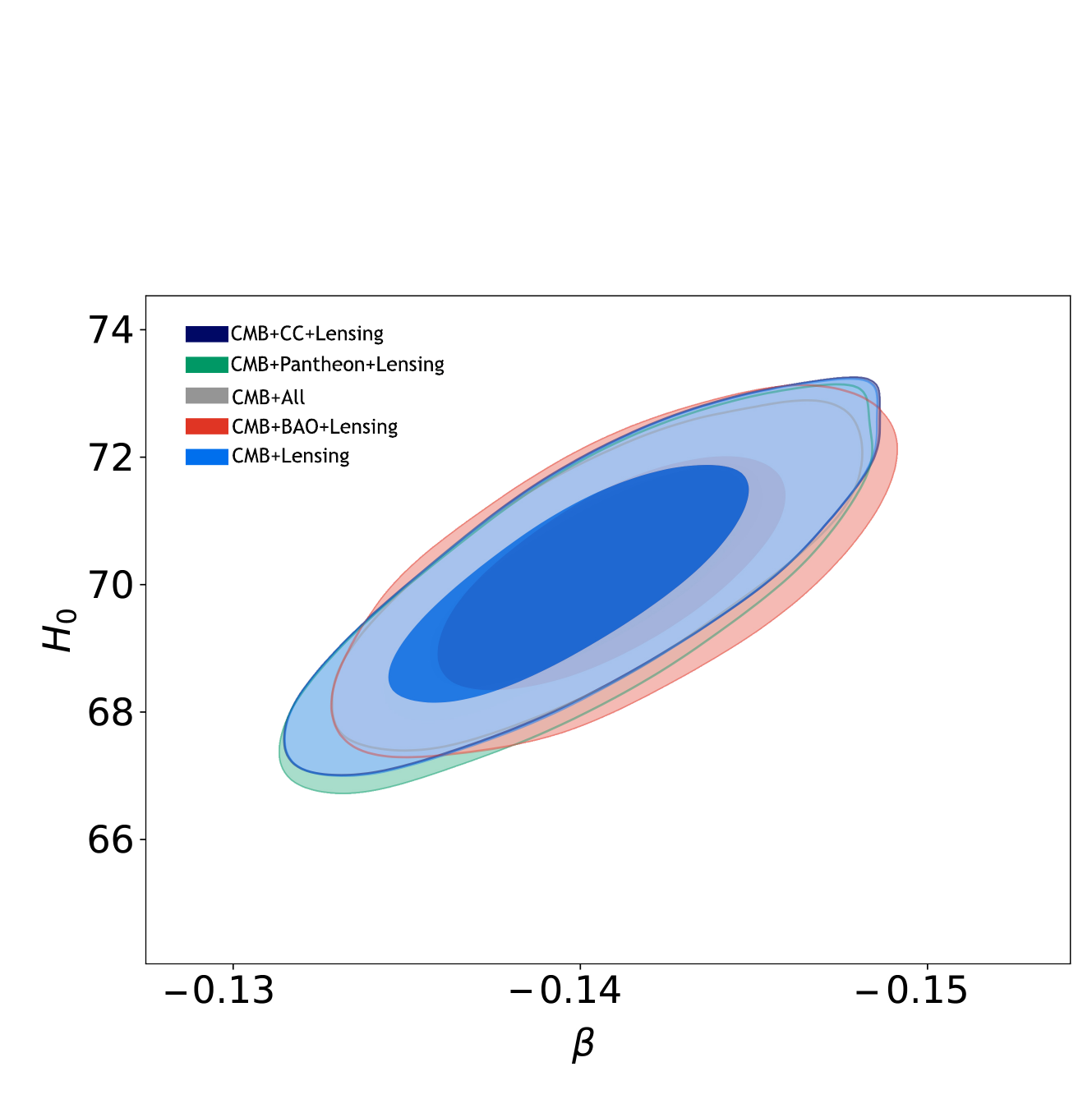}\hspace{0.1 cm}\\
	Fig. 11: The comparison of  $\beta$ parameter according to $H_{0}$ for different combination of datasets for  Interacting Barrow Holographic dark energy \\
\end{figure}

\begin{figure}
	\includegraphics[scale=.35]{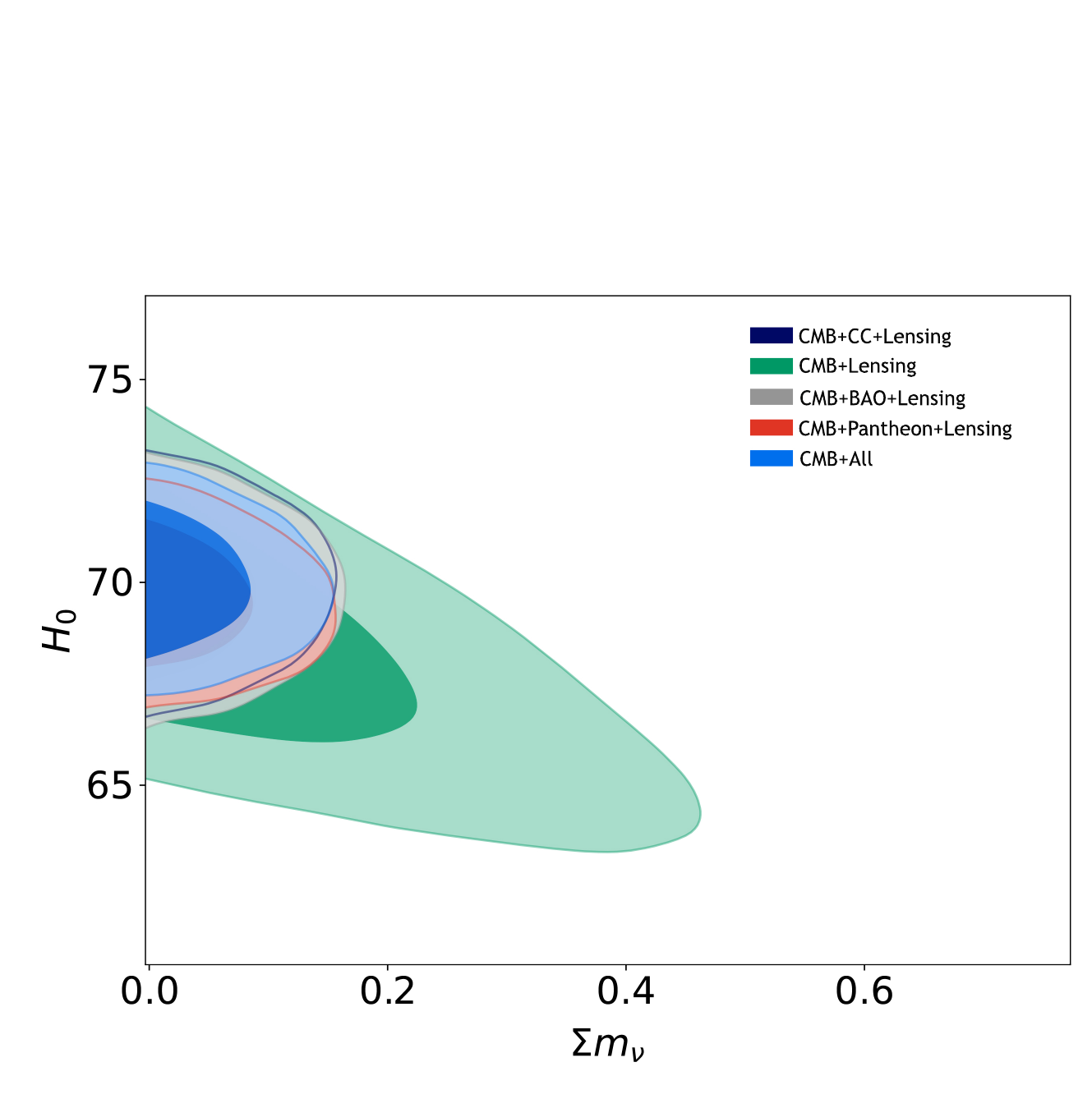}\hspace{0.1 cm}\\
	Fig. 12: Constraints at the (95$\% $CL.) two-dimensional contours for $\sum m_{\nu}$ for different combination of datasets in Interacting Barrow Holographic dark energy \\
\end{figure}

\begin{figure}
	\includegraphics[scale=.35]{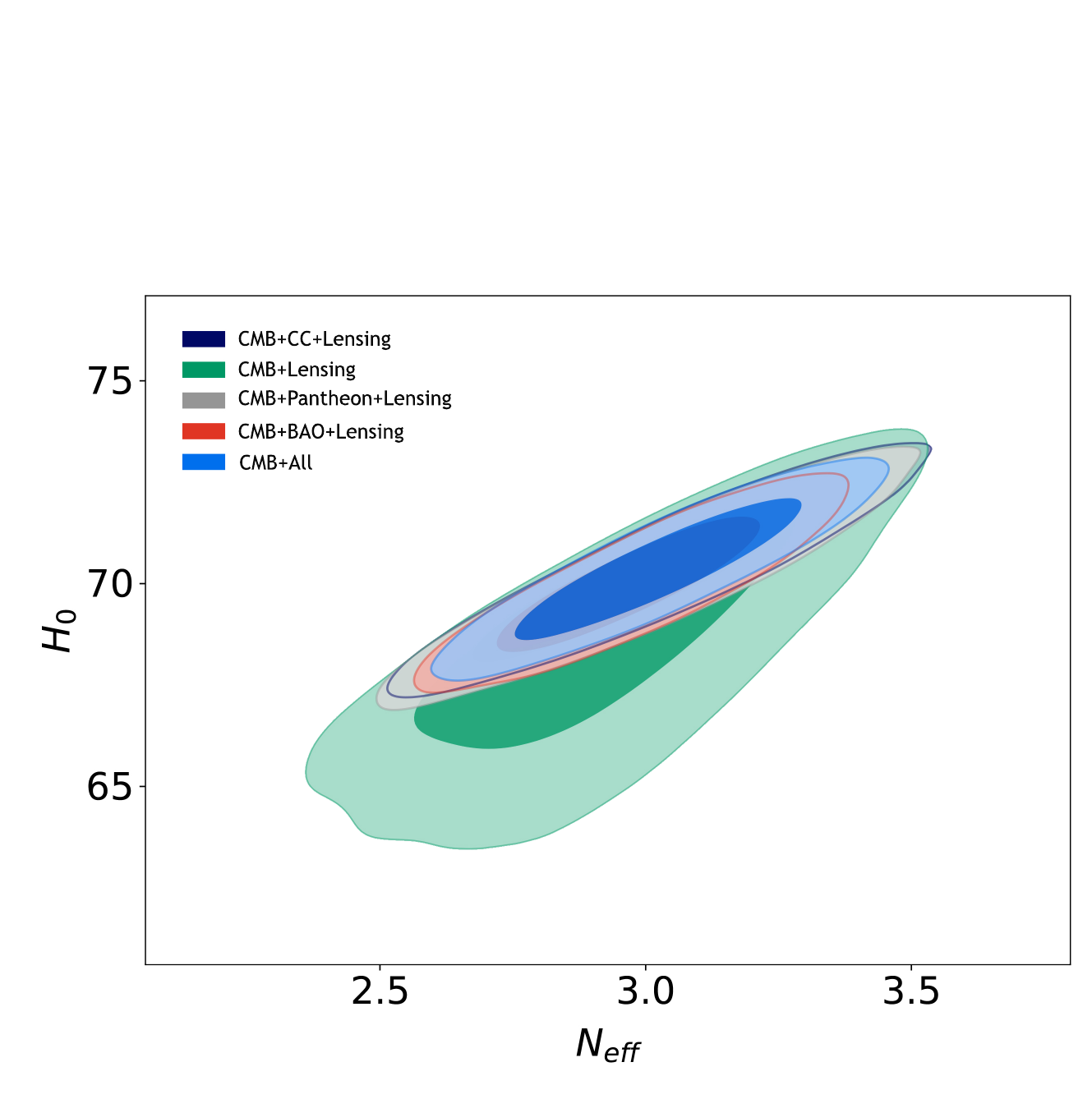}\hspace{0.1 cm}\\
	Fig. 13: Constraints at the (68$\% $CL.) two-dimensional contours for $N_{eff}$ for different combination of datasets in Interacting Barrow Holographic dark energy \\
\end{figure}

\begin{figure}
	\includegraphics[scale=.35]{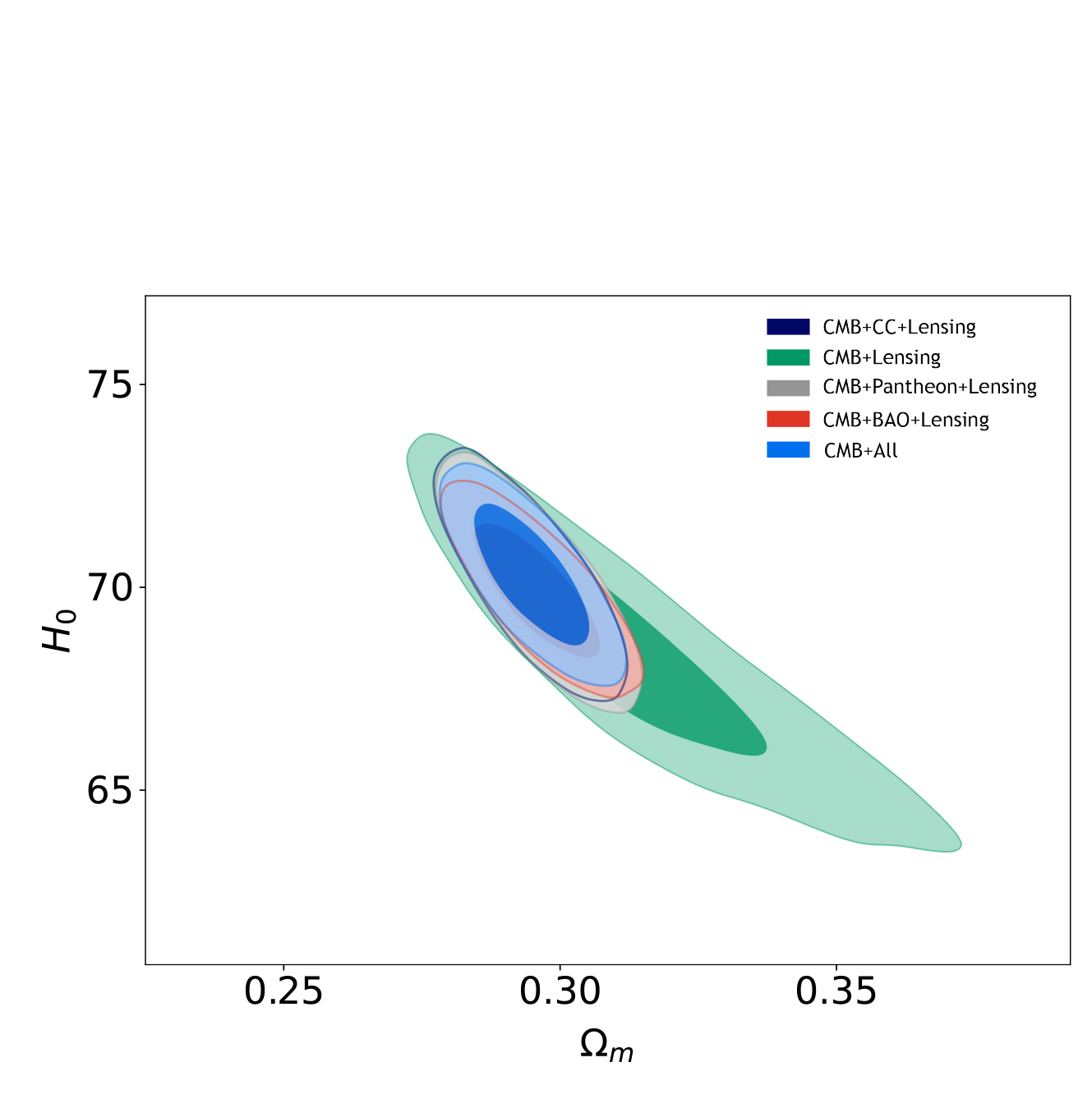}\hspace{0.1 cm}\\
	Fig. 14: Comparison  results of the  $\Omega_{m}$  according to $H_{0}$ for different combination datasets for Interacting Barrow Holographic dark energy \\
\end{figure}

\begin{figure}
	\includegraphics[scale=.35]{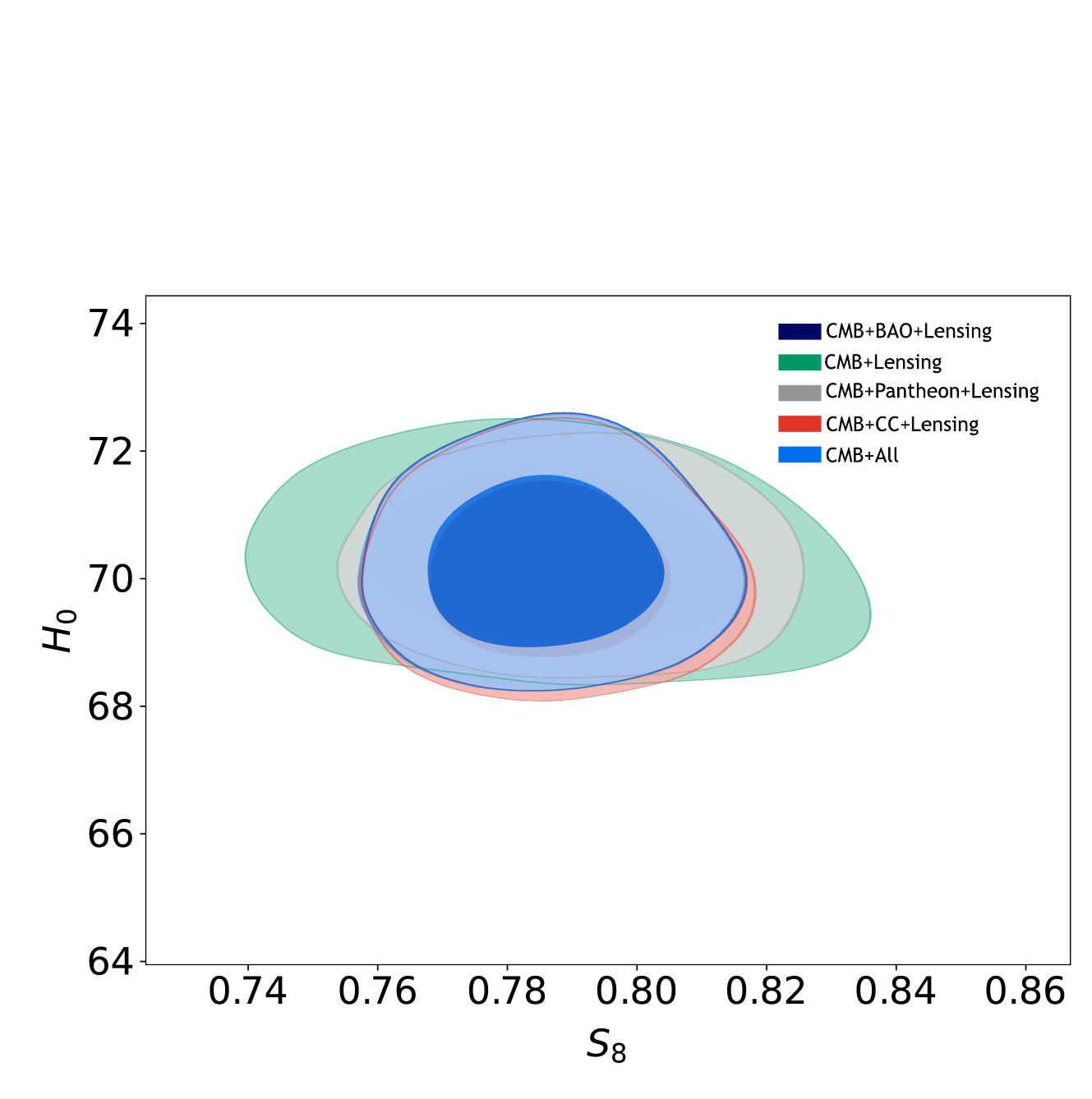}\hspace{0.1 cm}\\
	Fig. 15: Comparison  results of the  $S_{8}$ according to $H_{0}$ for different combination datasets for Interacting Barrow Holographic dark energy  \\
\end{figure}

\begin{figure*}
	\includegraphics[scale=.65]{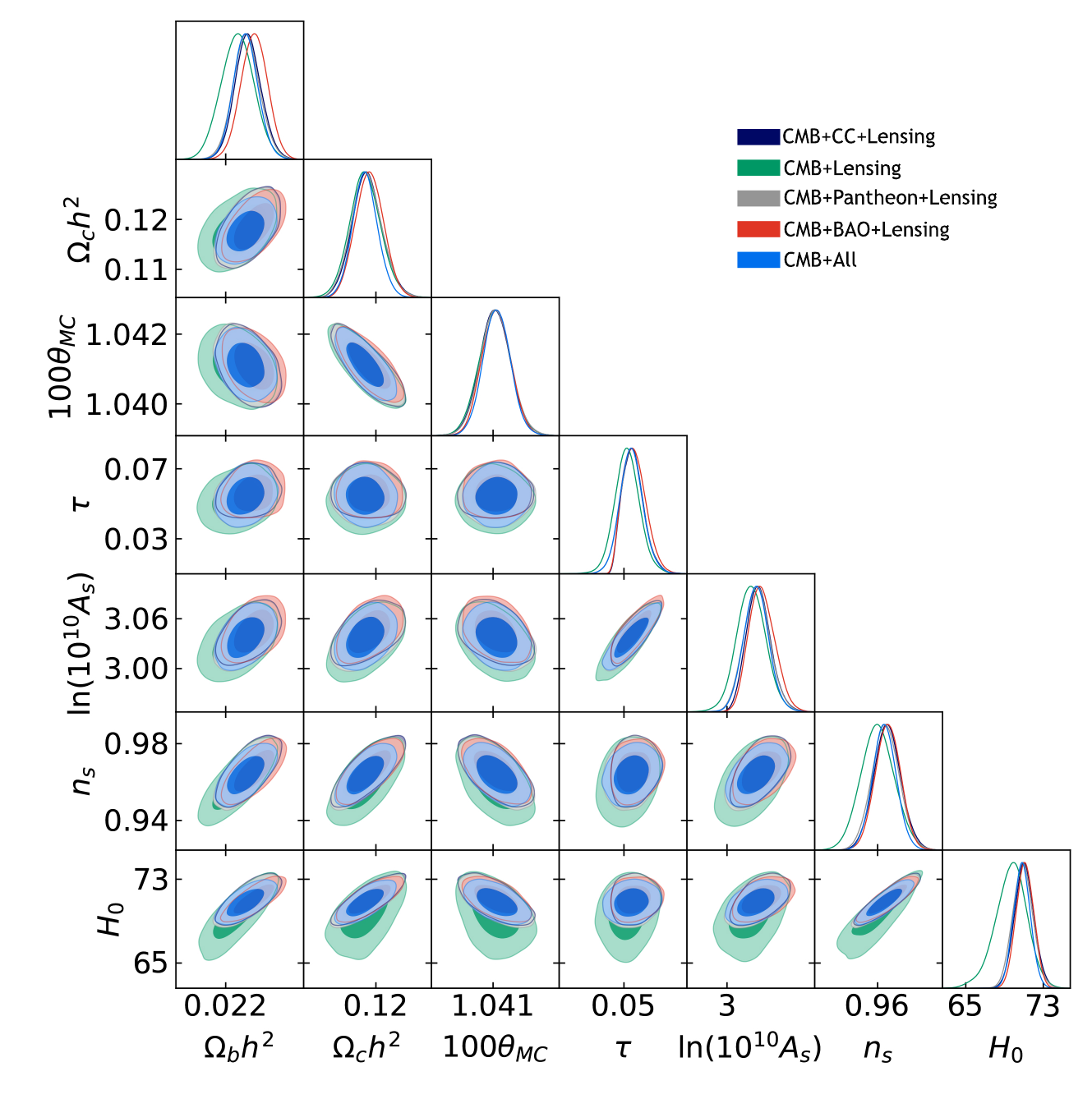}\hspace{0.1 cm}\\
	Fig. 16:  Comparison  results of the  $S_{8} $ , $\Omega_b h^2   $, $\Omega_c h^2   $, $n_s  $, ${\rm{ln}}(10^{10} A_s)$, $100\theta_{MC} $, $H_{0}(km/s/Mpc
	)$ for different combination datasets for Interacting Barrow Holographic dark energy\\
\end{figure*}

Figure 17 illustrates the comparative results of the matter clustering parameter (\( S_{8} \)) for various combinations of observational datasets within the Interacting Barrow Holographic Dark Energy model. This comparison underscores the influence of different dataset combinations on the derived \( S_{8} \) values, offering insights into the model's effectiveness in addressing the discrepancies related to the amplitude of matter fluctuations observed in large-scale structure surveys.
\begin{figure*}
	\includegraphics[scale=.53]{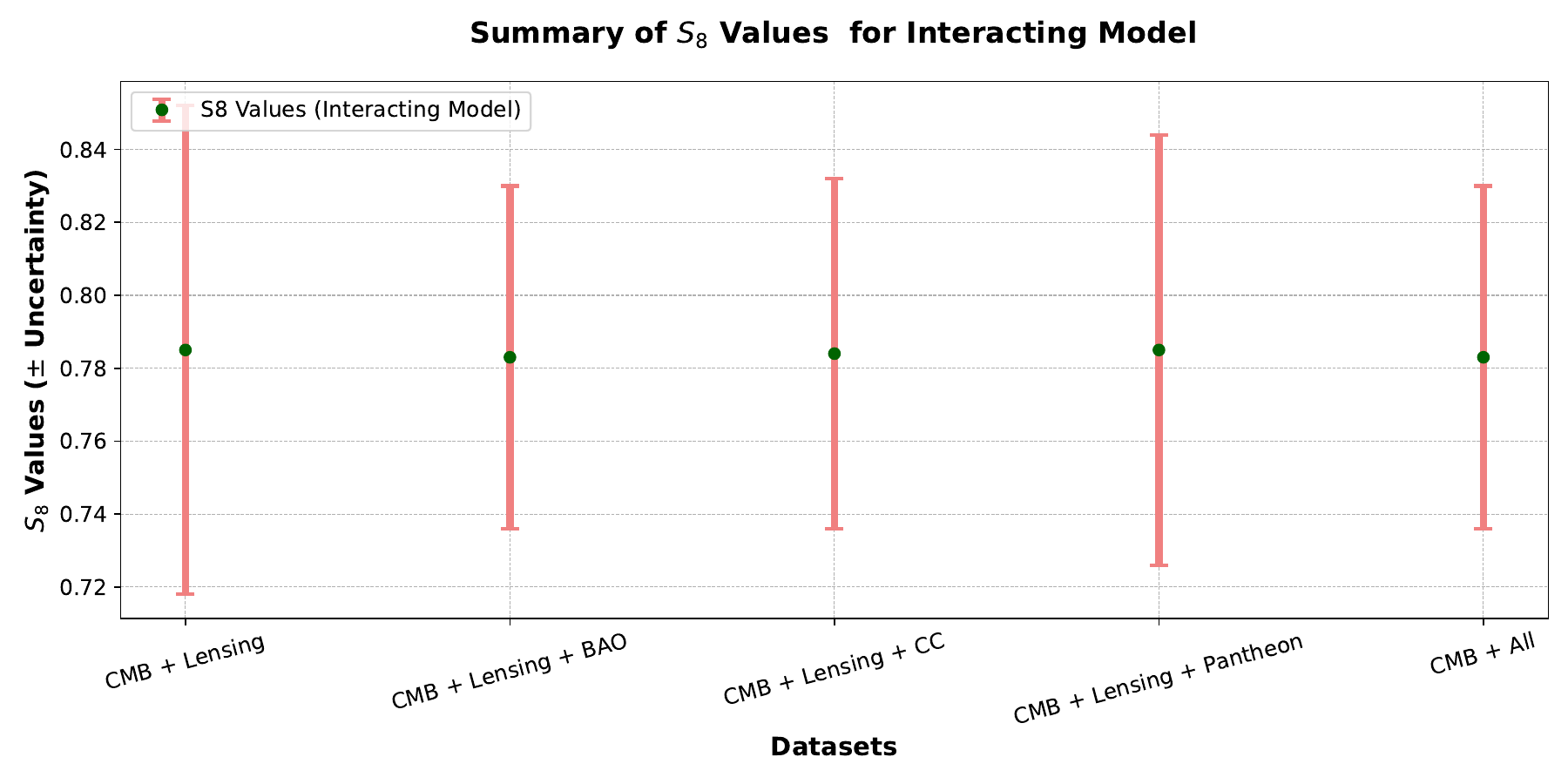}\hspace{0.1 cm}\\
	Fig. 17: Comparison  results of the  $S_{8}$  for different combination datasets for Interacting Barrow Holographic dark energy \\
\end{figure*}

The results presented in Table X demonstrate that the Interacting Barrow Holographic Dark Energy (BHDE) model provides a superior fit to the observational data compared to both the $\Lambda$CDM and Non-Interacting BHDE models. This conclusion is drawn from the consistently lower $\chi^2$ values across all dataset combinations explored in this work.

For the dataset combination CMB+lensing, the Interacting BHDE model achieves a total $\chi^2_{\rm tot}$ of $2780.331$, which is lower than both the $\Lambda$CDM model ($2787.123$) and the Non-Interacting BHDE model ($2785.452$). This trend continues across more complex dataset combinations, including CMB+CC+lensing, CMB+BAO+lensing, CMB+Pantheon+lensing, and CMB+all, where the Interacting BHDE model consistently yields lower total $\chi^2$ values. The superior performance of the Interacting BHDE model is particularly noticeable in the full dataset combination (CMB+all), where it achieves a $\chi^2_{\rm tot}$ of $3837.774$, compared to $3859.721$ for $\Lambda$CDM and $3849.881$ for Non-Interacting BHDE.

The improved fit of the Interacting BHDE model can be attributed to its ability to capture the interactions between dark energy and dark matter, which are not accounted for in the $\Lambda$CDM and Non-Interacting BHDE models. These interactions potentially provide a more accurate description of the underlying physics of the universe, leading to a better fit with the observational data.

Moreover, the lower $\chi^2$ values associated with the cosmic microwave background (CMB) and lensing datasets indicate that the Interacting BHDE model better accommodates the detailed structure of the early universe. This suggests that the model not only fits the overall cosmological observations more effectively but also aligns more closely with the specific features of the CMB, which is a crucial aspect of cosmological modeling.
\begin{table*}[t]   
	\caption{Observational constraints at $68\%$ on main and derived parameters of the Interacting scenario and Cosmological Parameter Results for Different Datasets. The parameter $\sum m_{\nu}$ reported at the $95\%$ CL, is in units of eV.}
	\centering
	\resizebox{1\textwidth}{!}{
		\begin{tabular}{|c|c|c|c|c|c|}
			\hline
			Parameter & CMB+lensing & CMB+lensing+BAO & CMB+lensing+CC & CMB+lensing+Pantheon & CMB+All \\
			\hline
			${\Omega_b h^2}$ & $0.02235 \pm 0.00021$ & $0.02233 \pm 0.00020$ & $0.02229 \pm 0.00024$ & $0.02228 \pm 0.00021$ & $0.02220 \pm 0.00021$ \\
			\hline
			${\Omega_c h^2}$ & $0.1193 \pm 0.0035$ & $0.1194 \pm 0.0033$ & $0.1191 \pm 0.0037$ & $0.1195 \pm 0.0034$ & $0.1192 \pm 0.0026$ \\
			\hline
			${100\theta_{MC}}$ & $1.04112 \pm 0.00046$ & $1.04115 \pm 0.00045$ & $1.04089 \pm 0.00056$ & $1.04090 \pm 0.00055$ & $1.04100 \pm 0.00050$ \\
			\hline
			${\tau}$ & $0.05525 \pm 0.0080$ & $0.0551 \pm 0.0082$ & $0.0544 \pm 0.0077$ & $0.0549 \pm 0.0077$ & $0.0552^{+0.0055}_{-0.0079}$ \\
			\hline
			${\mathrm{ln}(10^{10} A_s)}$ & $3.045 \pm 0.017$ & $3.043 \pm 0.017$ & $3.045 \pm 0.017$ & $3.046 \pm 0.016$ & $3.047^{+0.016}_{-0.017}$ \\
			\hline
			$n_s$ & $0.9646 \pm 0.0075$ & $0.9648 \pm 0.0076$ & $0.9673 \pm 0.0086$ & $0.9683 \pm 0.0084$ & $0.9657 \pm 0.0076$ \\
			\hline
			$\Delta$ & $<0.41$ & $<0.331$ & $<0.32$ & $<0.32$ & $<0.3$ \\
			\hline
			$\Omega_m$ & $0.312^{+0.024}_{-0.022}$ & $0.296^{+0.016}_{-0.013}$ & $0.295^{+0.017}_{-0.017}$ & $0.296^{+0.017}_{-0.018}$ & $0.296^{+0.012}_{-0.013}$ \\
			\hline
			$\sum m_{\nu} \ (\mathrm{eV})$ & $<0.32$ & $<0.124$ & $<0.124$ & $<0.123$ & $<0.122$ \\
			\hline
			$N_{eff}$ & $2.95^{+0.27}_{-0.32}$ & $2.97^{+0.18}_{-0.19}$ & $2.99^{+0.22}_{-0.22}$ & $2.97^{+0.23}_{-0.23}$ & $2.99^{+0.15}_{-0.14}$ \\
			\hline
			$\beta$ & $-0.138$ & $-0.142$ & $-0.14$ & $-0.141$ & $-0.139$ \\
			\hline
		\end{tabular}
	}
	\label{table:merged}
\end{table*}

\begin{table*}
	\caption{{\small $\chi^2_{}$s comparison between $\Lambda$CDM, Non-Interacting BHDE, and Interacting BHDE for the different dataset combinations explored in this work. CMB+all refers to Planck+lensing+BAO+CC+Pantheon.}}
	\begin{center}
		\resizebox{0.95\textwidth}{!}{  
			\begin{tabular}{ c |c c c c c  } 
				\hline
				\hline
				Model  & CMB+lensing & CMB+CC+lensing & CMB+BAO+lensing & CMB+Pantheon+lensing & CMB+all \\ 
				\hline
				$\Lambda$CDM  & $2787.123$ & $2797.504$ & $2785.82$ & $3818.997$ & $3859.721$  \\
				$\chi^2_{\rm  CMB}$ & $2778.132$ & $2768.123$ & $2771.070$ & $2767.707$ & $2779.518$  \\
				$\chi^2_{\rm lensing}$ & $8.991$ & $8.254$ & $8.174$ & $9.012$ & $9.520$  \\
				$\chi^2_{\rm  CC}$ & $-$ & $21.127$ & $-$ & $-$ & $21.424$  \\
				$\chi^2_{\rm  BAO}$ & $-$ & $-$ & $6.576$ & $-$ & $7.481$  \\
				$\chi^2_{\rm  Pantheon}$ & $-$ & $-$ & $-$ & $1042.278$ & $1041.778$  \\
				\hline
				Non-Interacting BHDE  & $2785.452$ & $2793.871$ & $2782.234$ & $3817.334$ & $3849.881$  \\
				$\chi^2_{\rm  CMB}$ & $2776.321$ & $2765.897$ & $2768.562$ & $2766.991$ & $2774.231$  \\
				$\chi^2_{\rm lensing}$ & $9.131$ & $8.621$ & $8.182$ & $9.005$ & $9.325$  \\
				$\chi^2_{\rm  CC}$ & $-$ & $19.353$ & $-$ & $-$ & $20.189$  \\
				$\chi^2_{\rm  BAO}$ & $-$ & $-$ & $5.490$ & $-$ & $6.371$  \\
				$\chi^2_{\rm  Pantheon}$ & $-$ & $-$ & $-$ & $1041.338$ & $1039.765$  \\
				\hline
				Interacting BHDE  & $2780.331$ & $2786.452$ & $2777.912$ & $3811.212$ & $3837.774$  \\
				$\chi^2_{\rm  CMB}$ & $2771.412$ & $2760.781$ & $2763.412$ & $2764.233$ & $2769.541$  \\
				$\chi^2_{\rm lensing}$ & $8.919$ & $7.621$ & $7.532$ & $8.012$ & $7.142$  \\
				$\chi^2_{\rm  CC}$ & $-$ & $16.952$ & $-$ & $-$ & $15.785$  \\
				$\chi^2_{\rm  BAO}$ & $-$ & $-$ & $4.821$ & $-$ & $4.995$  \\
				$\chi^2_{\rm  Pantheon}$ & $-$ & $-$ & $-$ & $1037.545$ & $1036.311$  \\
				
				\hline
				\hline
			\end{tabular}
		}
	\end{center}
	\label{table_chi_all_models}
\end{table*}

One of the noteworthy findings in our analysis is the persistent unconstrained nature of the parameter \( \Delta \), even when interactions between dark energy and matter are included. This lack of constraint suggests that the current observational data do not possess sufficient sensitivity to tightly bound the value of \( \Delta \) within the interacting model framework. 

Several factors may contribute to this result. Firstly, the precision of the employed datasets might not be adequate to distinguish between the interacting and non-interacting models effectively, leading to a broad allowed range for \( \Delta \). Secondly, the inherent insensitivity of the parameter itself could play a role, implying that \( \Delta \) may vary significantly without causing noticeable changes in observable quantities. Lastly, there might be a substantial correlation between \( \Delta \) and other cosmological parameters, causing variations in \( \Delta \) to be compensated by adjustments in other parameters, thus masking its direct influence on the model's predictions.

\section{Conclusion}

In this study, we have conducted a comprehensive analysis of the $H_0$ and $S_8$ tensions within the framework of both non-interacting and interacting Barrow Holographic Dark Energy (BHDE) models. By employing a combination of observational datasets, including CMB, BAO, cosmic chronometers (CC), Pantheon, and lensing data, we have provided a detailed assessment of the degree of tension in comparison to the widely acknowledged Planck 2018 results and recent measurements such as R22 for $H_0$, and KiDS-1000 and DES-Y3 for $S_8$.

Our findings reveal that both the non-interacting and interacting BHDE models contribute to alleviating the $H_0$ and $S_8$ tensions when compared to the standard $\Lambda$CDM model. The inclusion of all datasets in the non-interacting BHDE model yields a moderate tension with the Planck 2018 $H_0$ measurement, quantified at $1.52\sigma$. This tension, while significant, is reduced compared to the higher tension typically reported in the $\Lambda$CDM model, indicating that the interacting BHDE model offers a better fit to the $H_0$ data. The tension with the R22 measurement is slightly lower, at $1.45\sigma$, further suggesting that the non-interacting BHDE model is effective in reducing the discrepancy between local and CMB-inferred $H_0$ values.

For the interacting BHDE model, the $H_0$ tension with Planck 2018 is observed to be $1.43\sigma$, slightly lower than in the non-interacting scenario and still indicative of a reduction in tension compared to the $\Lambda$CDM model. The tension with R22, quantified at $1.13\sigma$, is also alleviated, demonstrating that the interacting BHDE model provides a viable approach to mitigating the $H_0$ tension.

Regarding the $S_8$ parameter, the non-interacting BHDE model shows a maximum tension of $1.28\sigma$ with Planck 2018 when all datasets are combined. This tension is lower than that typically found in the $\Lambda$CDM framework, indicating that the non-interacting BHDE model effectively alleviates the $S_8$ tension. The interacting BHDE model exhibits a maximum $S_8$ tension of $0.99\sigma$ with Planck 2018, further reducing the tension and suggesting that the interaction between dark energy and dark matter may play a role in better aligning the $S_8$ values with observational data.

When comparing our results with the KiDS-1000 and DES-Y3 surveys, we observe lower tensions across both models, with the interacting BHDE model generally showing better agreement with these weak lensing surveys. This suggests that the inclusion of interaction between dark energy and dark matter may offer a slight advantage in addressing the $S_8$ tension, although the overall impact is modest.

In conclusion, both the non-interacting and interacting BHDE models provide valuable insights into the $H_0$ and $S_8$ tensions, and both models contribute to alleviating these tensions when compared to the $\Lambda$CDM model. The non-interacting model demonstrates a somewhat better fit for $H_0$, particularly in relation to the R22 measurement, while the interacting model shows a slight edge in reconciling the $S_8$ parameter with weak lensing surveys. However, the tensions observed in both models underscore the ongoing challenges in fully resolving these cosmological tensions, pointing to the need for further theoretical developments and more precise observational data. The results presented here contribute to the ongoing discourse on the nature of dark energy and the challenges in achieving a consistent cosmological model that can reconcile the diverse range of observational datasets.

Also, our analysis, based on the CMB+All dataset, highlights the differences between the Interacting and Non-Interacting scenarios, both of which incorporate quantum-gravitational effects through the parameter $\Delta$. In the Interacting model, the sum of neutrino masses, $\sum m_{\nu}$, is constrained to be less than $0.122$ eV, which is slightly more stringent than the constraint in the Non-Interacting model. For the Non-Interacting scenario, the upper bound on $\sum m_{\nu}$ is also $<0.124$ eV.

The effective number of relativistic degrees of freedom, $N_{\rm eff}$, is similarly constrained in both models, with the Interacting scenario yielding $N_{\rm eff} = 2.99^{+0.15}_{-0.14}$, while the Non-Interacting scenario gives $N_{\rm eff} = 2.99^{+0.19}_{-0.2}$. These results suggest a comparable fit to the observational data for this parameter, with the Interacting model offering slightly tighter constraints.

The quantum-gravitational deformation parameter $\Delta$ is constrained to be less than 0.3 in the Interacting scenario and less than 0.43 in the Non-Interacting scenario. Since $\Delta$ must be non-negative due to its physical interpretation, these upper bounds indicate the extent to which quantum-gravitational effects influence each model. The tighter constraint on $\Delta$ in the Interacting model suggests that it may provide a more accurate representation of these effects.

In summary, while quantum-gravitational deformations are a feature of both models, the Interacting scenario, particularly when considering the CMB+All dataset, offers a slightly more robust fit to the observational data, making it a compelling alternative to the Non-Interacting scenario.


\section{Data availability statement}

The manuscript has no associated data or the data will not be deposited

\end{document}